\documentclass{article}
\usepackage[utf8]{inputenc}
\usepackage{amsmath}
\usepackage{graphics}
\usepackage{graphicx,color}
\usepackage{amsbsy,amssymb,amsmath,ulem}
\usepackage[all]{xy}
\usepackage{authblk}
\usepackage{comment}

\newcommand{\rr}{\mathbf{r}}

\newcommand{\diff}{\,\mathrm{d}^3}

\newcommand{\dive}{\mathrm{div}}

\newcommand{\xx}{\mathbf{x}}

\newcommand{\pp}{\mathbf{p}}

\newcommand{\eps}{\varepsilon}

\newcommand{\intd}[1]{\int\!\!\diff #1}

\newcommand{\cref}{c_0}

\newcommand{\stg}{{\ensuremath{-\kern-4pt{\ominus}\kern-4pt-}}}

\newcommand{\EE}{\mathbf{E}}
\newcommand{\BB}{\mathbf{B}}
\newcommand{\AAA}{\mathbf{A}}

\begin{document}
\title{Hamiltonian Coupling of Electromagnetic Field and Matter}

\author[1]{O\u{g}ul Esen}
\author[2,3,4]{Michal Pavelka\thanks{pavelka@karlin.mff.cuni.cz}}
\author[2]{Miroslav Grmela}

\affil[1]{Department of Mathematics, Gebze Technical University, Gebze-Kocaeli 41400, Turkey}
\affil[3]{New Technologies - Research Centre, University of West Bohemia,  Univerzitn\'{i} 8, 306 14 Pilsen, Czech Republic}
\affil[4]{Mathematical Institute, Faculty of Mathematics and Physics, Charles University in Prague, Sokolovsk\'{a} 83, 186 75 Prague, Czech Republic}
\affil[2]{École Polytechnique de Montréal, C.P.6079 suc. Centre-ville, Montréal, H3C 3A7, Québec, Canada}

\maketitle
\date

\begin{abstract}
Reversible part of evolution equations of physical systems is often generated by a Poisson bracket. We discuss geometric means of construction of Poisson brackets and their mutual coupling (direct, semidirect and matched pair products) as well as projections of Poisson brackets to less detailed Poisson brackets.
This way the Hamiltonian coupling of transport of mixtures with electrodynamics is elucidated.
\end{abstract}

\tableofcontents

\section{Introduction}

Dynamics has two main points of view, the Lagrangian dynamics and the Hamiltonian
dynamics \cite{Abraham-Marsden,Arnoldbook,libermann2012symplectic,Marsden-Ratiu}. The Lagrangian approach is based on the observation that there are variational
principles behind Newton's second law. In this approach, the dynamics of a system
is generated by a Lagrangian function on velocity phase space of the configuration
space. On the other hand, the Hamiltonian view of dynamics is based on symplectic geometry. In this approach, the dynamics is represented by a Hamiltonian function
on the momentum phase space. Transformations between the Lagrangian and the Hamiltonian dynamics are achieved by the Legendre transformations. If a non-degeneracy condition, called Hessian condition, is satisfied, then the transformation is immediate. Although there are some generalized versions of the Legendre transformation with no need of the non-degeneracy condition \cite{Di64,Tu77}, one usually faces serious complications when transforming different descriptions of the particular systems in the degenerate case.

In the literature the applications of the Hamiltonian and the Lagrangian dynamics diverse from the control theory  \cite{agrachev2013control,bloch1996nonholonomic,FB-ADL:04, jurdjevic1997geometric} to the image registration \cite{bruveris2011momentum,bruveris2015geometry}, and even to some DNA models \cite{ellis2010symmetry}. Additionally, strong motivation for studying geometrical structures of physical systems can be found in non-equilibrium thermodynamics as well. Many mesoscopic models of physical systems, e.g. kinetic theory, hydrodynamics, extended hydrodynamics of polymeric fluids and turbulence \cite{Miroslav-turbulence} or dynamics of plastic deformations \cite{hutter-plastic}  have been shown to posses the GENERIC structure \cite{GO,OG}, where the reversible evolution is expressed in Hamiltonian form while the irreversible evolution is given by a dissipation potential and entropy. Hamiltonian evolution can thus be coupled with irreversible gradient dynamics. An advantage of such coupling is that for example Onsager-Casimir reciprocal relations are satisfied automatically and generalized into the far-from-equilibrium regime \cite{PRE2014}. Another advantage is that one can use the geometrical results developed in theory of Hamiltonian dynamics \cite{marsden1984reduction}.

Formulating reversible part of evolution in the Hamiltonian sense is a modern approach in non-equilibrium thermodynamics. Is such an approach compatible with other results of non-equilibrium thermodynamics, such as the second law of thermodynamics and Onsager-Casimir reciprocal relations? The Hamiltonian evolution is compatible with the second law of thermodynamics in the sense that total entropy (on a chosen particular level of description) of an isolated system is not changed by the evolution. The growth of entropy is thus realized only within the irreversible evolution, for example gradient dynamics, which is not Hamiltonian.

Onsager-Casimir reciprocal relations, which can be seen as restrictions on how variables can be coupled consistently, are also often fulfilled, see \cite{HCO,PRE2014}. It is shown at the end of this paper in which sense the Hamiltonian coupling fulfills the Onsager-Casimir reciprocal relations. Therefore, the Hamiltonian coupling between electromagnetic field and matter is compatible with non-equilibrium thermodynamics.

In the present paper we consider only the reversible parts of the evolution equations, and the Hamiltonian representations of the dynamical systems. The canonical representation of the Hamiltonian dynamics can be formulated on a symplectic manifold which is, by definition, even dimensional. Hence odd dimensional and some infinite dimensional systems, such as rigid bodies, thermodynamics, fluid and plasma theories, can not be expressed in the framework of canonical Hamiltonian formalism. For such systems, one may consult the Hamiltonian reduction theory which proposes some methods to obtain a non-canonical Hamiltonian formulation from a canonical one by dividing out the symmetries or/and the constraints. Most of the cases, the reduction procedure results in a Poisson structure. Poisson geometry is a generalization of the symplectic geometry and obtained technically by relaxing the non-degeneracy requirement of the symplectic two-form. Although the origins of Hamiltonian reduction theory can be found in the works of Euler, Lagrange, Hamilton, Jacobi and Poincar\'{e}, the start of the modern history of the geometrization theory can be considered as the pioneering papers of Arnold \cite{Arnold} and Smale \cite{smale1970topology}. The geometrization of the Hamiltonian reduction theory achieved by Marsden and Weinstein \cite{marsden1974reduction}, see also \cite{meyer1973symmetries}. We, additionally, refer to \cite{marsden2007hamiltonian} for a brief history of this theory.

Let us now present the motivation of the present work on the abstract level. Consider two Hamiltonian systems in mutual interactions. It is evident that in their collective motion the constitutive systems cannot keep their individual motions due to the presence of the mutual interactions. In other words, the equations governing coupled (matched) systems cannot be obtained merely by putting together the individual equations of the constitutive systems. The equations of motion of the coupled (matched)  system may be obtained by adding extra terms to the individual equations of motions, and these extra terms are determined by the geometries of the constitutive systems as well as by the forms of the interactions. Note that while performing a coupling, it is not necessary to have two different systems. One can instead impose coupling of two faces of a single dynamical system. For example, to determine the behavior of a fluid with its electromagnetic properties, we write magnetoelectrohydrodynamic equations, which are the coupling of Maxwell and Euler equations.

The main problem addressed in this present paper is to determine the matched equations of motion of two interacting systems (whose configuration spaces are Lie groups) governing the coupled system starting from the individual equations of motions. The Lie-group characterization of the configuration spaces of the systems is imperative here to define the mutual actions. The geometrical construction we propose does not have any particular restrictions, and it can be used for any two systems in mutual interaction satisfying certain compatibility conditions \cite{esen2015lagrangian,esen2016hamiltonian}. The matched pair concept that we shall present is the most general geometric way of coupling two systems in mutual interactions. It is a generalization of the semi-direct product theory, where only one of the constitutive system acts on the other. We remark here also that knowing how to couple (match) two systems leads to a deep understanding of how to decouple a system into two of its subsystems. So, if one achieves to write a system as a matched pair, then, applying the theory we are presenting, the system can be decoupled into two of its subsystems in a purely geometrical way.

The novelty of this paper lies in the following points. First of all, we shall fill the gap of the application of the matched pair technique in the case of field theories. Further, this is the first time to apply the matched pairs for physical systems whose configuration spaces are infinite dimensional. In particular, we shall present Hamiltonian formulations of various different couplings of electromagnetic field and matter, namely kinetic electrodynamics, magnetohydrodynamics, electromagnetohydrodynamics and their binary versions (describing binary mixtures). Although all the geometric frameworks of these particular physical examples have been presented in some previous pioneering studies, we shall collect them and try to show how a more general geometric framework (matched dynamics, c.f. Section (\ref{CTHS})) can be defined covering all these geometries. Secondly, the various Poisson brackets coupling matter with electromagnetic fields are identified as particular realizations of a hierarchy of Poisson brackets \cite{PhysicaD-2015}. Such a hierarchy makes the derivation easier and more accessible and gives also clearer physical meaning to the brackets. Although the brackets themselves can not be considered new, as they have been derived by means of Lie-Poisson reduction and related techniques, we believe that showing relations among the brackets together with derivation more accessible to physicists and engineers is worth mentioning. In this respect, we shall try to be very gentle while introducing and presenting the results in order to make this work more accessible for broader audience. Thirdly, Onsager-Casimir reciprocal relations implied by the Poisson brackets will be discussed. Such a discussion could help understanding how the reciprocal relations appear in mesoscopic evolution on different levels of description.

In accordance with the goal of making the paper as accessible as possible, we shall first recall the definitions of Hamiltonian systems, Poisson and Lie-Poisson structures in the following section. We shall present the Hamiltonian formulations of the Maxwell equations and of the reversible part of the Boltzmann equation. In the third section, we shall focus on the couplings of two systems in mutual interactions. In that case one defines a matched pair (or a bicross) product of the constitutive systems. In section four the Poisson brackets related with hydrodynamics, binary hydrodynamics and classical binary hydrodynamics will be presented. The fifth section is reserved for the couplings of hydrodynamics and plasma with electromagnetic (EM) field. In this respect, we shall point out two particular cases of the matched pair dynamics. Firstly, configuration space of the coupled system will be taken simply as the Cartesian (direct) product of the configuration spaces of constitutive systems. The kinetic electrodynamics and binary couplings are of this kind. Secondly, recalling the semidirect product, only one of the constitutive system acts on the other. Hydrodynamics, magnetohydrodynamics and electromagnetohydrodynamics are of this kind. The last section is reserved for the discussions on Onsager-Casimir reciprocal relations.

\section{Hamiltonian Systems}
\subsection{Poisson Structures}

The configuration space of a dynamical (mechanical) system can roughly be defined as the set of all possible states (positions) of the system. Configuration spaces are usually only locally Euclidean. That is, although one has a local coordinate frame at every instance, one can not find any global coordinate chart covering the whole domain. This is even true for the simplest systems. Consider, for example, the simple pendulum. Its configuration space is a circle, which cannot be covered by a single coordinate chart due to some topological obstructions \cite{nakahara2003geometry,spivak1981comprehensive,Fecko}. In general, a configuration space is an abstract geometrical object called manifold. Manifolds look locally like Euclidean spaces as desired and they are additionally equipped with coordinate transformations satisfying some compatibility conditions. For finite cases, dimension of a manifold is defined as the dimension of its local picture.

The cotangent bundle $T^{\ast }\mathcal{M}$ of a manifold $\mathcal{M}$ is itself a manifold consisting of positions and momenta \cite{abraham2012manifolds,cartan2012differential,suhubi2013exterior}. So, if $\mathcal{M}$ is $n$-dimensional with coordinates $(\mathbf{r})$ then $T^{\ast }\mathcal{M}$ is $2n$-dimensional with induced (Darboux') coordinates $\left(\mathbf{r},\mathbf{p}\right) $ representing the momenta $(\mathbf{p})$ in addition to the positions $(\mathbf{r})$. A Hamiltonian function $H$ is a real valued function defined on the cotangent bundle. In the classical dynamics, $H$ is interpreted as the total energy. Once a Hamiltonian function $H$ is chosen, the dynamics is described by the Hamiltonian vector field
\begin{equation} \label{Hvf}
X_{H}=\frac{\partial H}{\partial \mathbf{p}}\cdot\frac{\partial }{\partial \mathbf{r}} -\frac{\partial H}{\partial \mathbf{r}}\cdot \frac{\partial }{\partial \mathbf{p}}.
\end{equation}
The equations of motion along the Hamiltonian vector field are called the Hamilton's equations and can be written as 
\begin{equation}
\dot{\mathbf{r}}=\frac{\partial H}{\partial \mathbf{p}},\qquad \dot{\mathbf{p}}=-\frac{\partial H}{\partial \mathbf{r}}.  \label{Hamiltonianeq}
\end{equation}
Components of the Hamiltonian vector field are thus right hand sides of evolution equations of the respective state variables.

It is evident that the Hamilton's equations (\ref{Hamiltonianeq}) depend on the local coordinates. In order to write the equations in a coordinate free form, one uses the symplectic two-form $\Omega _{T^{\ast }\mathcal{M}}$ on the cotangent bundle $T^{\ast }\mathcal{M}$. $\Omega _{T^{\ast }\mathcal{M}}$ is a canonical two-form, which is closed and non-degenerate \cite{Abraham-Marsden,Arnold,libermann2012symplectic,Fecko}. In this symplectic framework,  the Hamiltonian vector field corresponding to a given Hamilton function is defined as
\begin{equation}
i_{X_{H}}\left( \Omega _{T^{\ast }\mathcal{M}}\right) =dH, \qquad \text{or}
\qquad \Omega _{T^{\ast }\mathcal{M}} (X_{H},Z)=dH\cdot Z, \forall Z \label{modham}
\end{equation}%
where $i_{X_{H}}$ is the contraction. The non-degeneracy of the symplectic two-form $\Omega _{T^{\ast }\mathcal{M}}$ guaranties the uniqueness of the Hamiltonian vector field for a Hamiltonian function modulo constants.

By taking the directional derivative of a function $F$ (defined on $T^*\mathcal{M}$) in the direction of the Hamiltonian vector field $X_H$, we arrive at the Poisson bracket
\begin{equation} \label{canPoisson-}
\left\{ F,H\right\}:=X_{H}\left( F\right)=\frac{\partial H}{\partial \mathbf{r}}\cdot \frac{\partial F}{\partial \mathbf{p}}-\frac{\partial H}{\partial \mathbf{p}}\cdot\frac{\partial F}{\partial \mathbf{r}}
\end{equation}
of two functions $F$ and $H$. It can be observed immediately that this definition of Poisson bracket holds for any two smooth functions, hence it is a well-defined operation on the space $\mathcal{F}(T^{\ast }\mathcal{M})$ of functions on $T^{\ast }\mathcal{M}$. The bracket in (\ref{canPoisson-}) is called the canonical Poisson bracket as it is defined by the canonical symplectic two-form $\Omega _{T^{\ast }\mathcal{M}}$. In this picture, the Hamilton's equations (\ref{Hamiltonianeq}) can be written as
\begin{equation}
\dot{\mathbf{r}}=\left\{ \mathbf{r},H\right\} \text{ \ \ and \ \ }\dot{\mathbf{p}}=\left\{
\mathbf{p},H\right\} .  \label{coham}
\end{equation}%
hence the evolution of the state variables become
\begin{eqnarray}
 \dot{F} &=& \frac{\partial F}{\partial \mathbf{r}} \cdot \dot{\mathbf{r}}+ \frac{\partial F}{\partial \mathbf{p}} \cdot \dot{\mathbf{p}} = \frac{\partial F}{\partial \mathbf{r}} \cdot \left\{ \mathbf{r},H\right\} + \frac{\partial F}{\partial \mathbf{p}} \cdot \left\{
\mathbf{p},H\right\} \notag \\
 &=& \frac{\partial F}{\partial \mathbf{r}} \cdot \frac{\partial H}{\partial \mathbf{p}} +
 \frac{\partial F}{\partial \mathbf{p}}   \cdot (-\frac{\partial H}{\partial \mathbf{r}})
 = \{F,H\}.
\end{eqnarray}
Note that we have defined three different, but equivalent, realizations of the Hamilton's equations given by Eqs.(\ref{Hamiltonianeq}), Eqs.(\ref{modham}) and Eqs.(\ref{coham}).

It is possible to define Poisson structures without referring a symplectic structure \cite{vaisman2012lectures}. A Poisson structure on a manifold $\mathcal{P}$ is a bilinear skew-symmetric binary operation $%
\left\{ \bullet,\bullet\right\} $ on the space $\mathcal{F}\left( \mathcal{P}\right)$ of smooth functions that satisfies
\begin{enumerate}

\item Jacobi identity: $\left\{ F_1,\left\{ F_2,F_3\right\} \right\} +\left\{
F_2,\left\{ F_3,F_1\right\} \right\} +\left\{ F_3,\left\{ F_1,F_2\right\} \right\} =0,$

\item Leibniz identity: $\left\{F_1 F_2,F_3\right\} =F_1\left\{ F_2,F_3\right\} +\left\{
F_1,F_3\right\} F_2$
\end{enumerate}
for all $F_1,F_2,F_3$ in $\mathcal{F}(\mathcal{P})$, and we define the associated Hamiltonian vector field $X_{H}$ by
\begin{equation} \label{canPoisson-2}
X_{H}\left( F\right):=\left\{ F,H\right\}.
\end{equation}
Note that since a non-degeneracy condition is not assumed, the dynamics $X_{H}$ for a given function $H$ may not be unique. In other words, the degeneracy, if any, brings ambiguity in the choice of Hamiltonian function $H$. The kernel of Poisson bracket may be non-trivial. If the kernel is non-trivial, then there exists non-constant functions $C$ called Casimir functions satisfying $\{F,C\}=0$, for all functions $F$ \cite{weinstein1983local}. It is easy to deduce from the definition in (\ref{canPoisson-2}) that, if $X_{H}$ is the Hamiltonian vector field for a Hamiltonian function $H$, then it is also Hamiltonian vector field for function $H+C$ where $C$ being a Casimir. In other terms, we have that $X_{H+C}=X_{H}$. This shows that Casimir functions are conserved under the Hamiltonian flow. This may be useful in the applications see, for example, \cite{van2004port}. Locally, a Poisson manifold is the union of symplectic leaves hence it is possible to define a set of coordinates $(\mathbf{r,p,w})$ where the functional structure of the Poisson bracket is the same with the canonical Poisson bracket presented in (\ref{canPoisson-}) except some additional coordinates $(\mathbf{w})$ \cite{weinstein1983local}. If a function depends only on $(\mathbf{w})$, it is immediate to observe that its Poisson bracket is zero for all the other functions hence it is a Casimir function. 

In the framework of the non-equilibrium thermodynamics, one of the Casimir functions is the entropy on the level of description where the Poisson bracket generates reversible evolution \cite{GO,OG}. This comes from the assumption that time-irreversible evolution (in the sense of time reversal transformation \cite{PRE2014}) is the dissipative evolution, whereas the entropy grows, while the time-reversible evolution does not change the total entropy. 

The differentiable transformations preserving the Poisson structures are of great importance for the present manuscript. A differentiable mapping $\varphi$ from a Poisson manifold $\left( \mathcal{P}_{1},\left\{
\bullet,\bullet\right\} _{1}\right) $ to another Poisson manifold $\left( \mathcal{P}_{2},\left\{ \bullet,\bullet\right\}
_{2}\right)$ is called a Poisson mapping if it respects the brackets, that is if 
\begin{equation}
\left\{ F,H\right\} _{2}\circ \varphi =\left\{ F\circ \varphi ,H\circ
\varphi \right\} _{1},
\end{equation}%
for all $F,H\in \mathcal{F}\left( \mathcal{P}_{2}\right)$.

\subsection{Electrodynamics} \label{sec.ED}

We present the Maxwell's equations in the canonical Hamiltonian form by following \cite{Marsden1982}. Let $\mathfrak{U}$ be the space of one-form sections (or literally vector potentials) on $\mathbb{R}^3$. After fixing a top-form (volume) $d\rr$ on $\mathbb{R}^3$, the space of momenta $T^\ast\mathfrak{U}=\mathfrak{U}\times \mathfrak{U}^*$ consists of two-tuples $(\mathbf{A},\mathbf{Y})$ both of which can be identified the vector fields on $\mathbb{R}^3$ as well. In the geometric field theories, the pairings between dual spaces are defined in terms of integrals. The canonical Poisson bracket on $T^\ast\mathfrak{U}$ is given by
\begin{equation} \label{eq.PB.ED.Can}
 \{F,H\}^{(EMc)}=  \intd\rr \frac{1}{\eps_0} (F_\mathbf{A} \cdot H_\mathbf{Y} - H_\mathbf{A} \cdot  F_\mathbf{Y}),
\end{equation}
where $F_\mathbf{A}$ is the functional derivative of the $F$ with respect to $\mathbf{A}$ and it is defined as
 \begin{equation}\label{FD}
   \langle F_\mathbf{A}, \delta \mathbf{A} \rangle = \frac{d}{d\epsilon}|_{\epsilon=0} F(\mathbf{A}+\epsilon \delta \mathbf{A}), 
 \end{equation}
 and $\eps_0$ is permittivity of vacuum.
 Technically speaking, although we are identifying the vectors and covectors, $F_\mathbf{A}$ is actually an element of the dual space $\mathfrak{U}^*$ due to assumed reflexivity.
 To perform the operation on the right hand side of the bracket (\ref{eq.PB.ED.Can}), one simply takes the dot product and then integrates.
 One may compare the canonical Poisson bracket (\ref{canPoisson-}) presented for the finite dimensional cases and the one in (\ref{eq.PB.ED.Can}) for the infinite dimensional $T^\ast\mathfrak{U}$. In the latter one, the pairing are given by integrations, and the partial derivatives are replaced by functional derivatives.

 We make the substitutions $\BB=\nabla \times \mathbf{A}^\sharp $ and $\EE=-\mathbf{Y}$, where $\mathbf{A}^\sharp\cdot \partial_\rr=\delta^i_j A_i \partial_j$ is the vector field constructed by the components of the one-form $\mathbf{A}\cdot d\rr =A_i dr^i$. The bracket (\ref{eq.PB.ED.Can}) then becomes
\begin{equation} \label{eq.PB.ED}
 \{F,H\}^{(EM)} =
 \intd\rr \frac{1}{\eps_0} \left(F_\EE \cdot (\nabla \times H_\BB) - H_\EE \cdot (\nabla \times F_\BB)\right),
\end{equation}
which governs the evolution of electromagnetic fields $\EE$ and $\BB$. With Hamiltonian function
\begin{equation*}
  H=(1/2)\intd\rr \eps_0(E^2 + c^2 B^2)
\end{equation*}
the Hamilton's equations are two of the Maxwell's equations
\begin{subequations}
\begin{eqnarray}\label{eq.Maxwell}
 \dot{\EE} &=& c^2 \nabla \times \BB, \\
 \dot{\BB} &=& -\nabla \times \EE,
 \end{eqnarray}
\end{subequations}
and the remaining two are the results of the gauge invariance. That is, the bracket (\ref{eq.PB.ED})
is endowed with the following constraints
\begin{subequations}  \label{Eq.Cons.ED}
\begin{eqnarray}
 \dive \EE &=& \frac{z e}{\eps_0}\rho, \\
 \dive \BB &=& 0,
\end{eqnarray}
\end{subequations}
where  $z$ is number of elementary charges per particle and $e$ is the elementary charge. Note that dots in Eqs. \eqref{eq.Maxwell} stand for partial derivatives with respect to time.

\subsection{The Lie-Poisson Formulation}

A symmetry of a differential equation is a transformation of dependent or/and independent variables preserving the structure of the equation, see e.g. \cite{olver2000applications}. The composition of two symmetry transformation is another symmetry of the system, that is the set of symmetries (transformations) is closed under the composition operation. This set is called a group if it additionally satisfies associativity and invertibility conditions.

A Lie group is a manifold $G$ that has a group structure
consistent with its manifold structure, in the sense that, group
multiplication and inversion
$$ G\times G \rightarrow G: (g,h) \rightarrow gh, \qquad G\rightarrow G: g\rightarrow g^{-1}  $$
are smooth maps. The particular case in which the configuration space of a dynamical system
is a Lie group attracts deep interest, since the configuration
spaces of the systems such as rigid body dynamics, fluid and
plasma theories, are Lie groups \cite{arnold1999topological}.

At the tangent space $\mathfrak{g}:=T_e G$ over the identity element $e$ of $G$, by taking the derivative of the inner automorphism \cite{Fecko} on the group $G$, one arrives at an anti-commutative algebra
\begin{equation} \label{LieBra}
  \mathfrak{g} \times \mathfrak{g} \rightarrow \mathfrak{g}: (\xi,\eta)\rightarrow [\xi,\eta]
\end{equation}
satisfying the Jacobi identity
\begin{equation}
\left[ \xi ,\left[ \eta ,\zeta \right] \right] +\left[ \eta ,\left[ \zeta
,\xi \right] \right] +\left[ \zeta ,\left[ \xi ,\eta \right] \right] =0,
\end{equation}%
$\forall \xi ,\eta ,\zeta \in \mathfrak{g}$. The bracket is called the Lie algebra bracket and two tuple $\left( \mathfrak{g},\left[\bullet ,\bullet\right] \right)$  is called a Lie algebra. The Lie algebra bracket defines the adjoint representation of $\mathfrak{g}$ on itself given by $$ ad_\xi: \mathfrak{g} \rightarrow \mathfrak{g}: \eta \rightarrow [\xi,\eta].$$

Existence of the Lie algebra structure leads to the definition of a Poisson bracket on the linear algebraic dual $\mathfrak{g}^*$ of $\mathfrak{g}$ called the Lie-Poison bracket  \cite{Marsden-Ratiu}. Explicitly, the Lie-Poisson bracket is defined by
\begin{equation} \label{LPBr}
\left\{ F,H\right\} \left( \mu \right) =   \langle \mu ,\left[ F_\mu ,H_ \mu \right]\rangle ,
\end{equation}
where $\mu  \in \mathfrak{g}^{\ast }$, $\left[ \bullet,\bullet\right] $
is the Lie bracket on $\mathfrak{g}$, and $\left\langle \bullet,\bullet
\right\rangle $ is the pairing between Lie algebra and its dual. For finite dimensional case $F_\mu$ stands for the partial derivative of the function whereas for the infinite dimensional cases, $F_\mu$ stands for the functional derivative of $F$ with respect to $\mu$ which is defined as
 $$ \langle F_\mu, \delta \mu \rangle = \frac{d}{d\epsilon}|_{\epsilon=0} F(\mu+\epsilon \delta \mu) $$
for all $\delta \mu \in \mathfrak{g}^*$. Compare this general definition and the one presented in (\ref{FD}). To make the Lie-Poisson bracket well-defined, we need additionally to assume that $F_\mu \in \mathfrak{g}^{\ast \ast }\simeq \mathfrak{g}$, i.e. the second dual of the Lie algebra is isomorphic to the algebra itself. To arrive at the
Hamilton's equations in this reduced picture, we compute the Hamiltonian vector field
\begin{eqnarray} \label{LPB}
X_{H}\left( F\right)= \left\{ F,H\right\}=\left\langle \mu  ,\left[
F_\mu,H_\mu\right]
\right\rangle 
= \langle \mu  ,ad_{F_\mu}H_\mu \rangle
= - \langle ad_{ H_\mu }^{\ast
}\mu  , F_\mu \rangle.
\end{eqnarray}
Here, the coadjoint action $ad^\ast _\xi$ is minus of the linear algebraic dual of adjoint action that is the Lie bracket $ad_\xi (\bullet)=[\xi,\bullet]$. The calculation shows that dynamics generated by a reduced Hamiltonian $H$ on the dual space is governed by the Lie-Poisson equations
\begin{equation}\label{LP}
  \dot{\mu} = - ad^\ast _{H_\mu}\mu.
\end{equation}

It is evident that one may multiply the Lie-Poisson bracket (\ref{LPBr}) by a minus sign without disturbing its functional analytic and algebraic properties. In this case, one needs to replace the minus sign in front of the Lie-Poisson equations with a plus. The sing in front of the Lie-Poisson bracket or/and Lie-Poisson equations is a manifestation of the (left/right) symmetry that the unreduced total system has. For the present paper, we are using the plus Lie-Poisson bracket (\ref{LPBr}) since the kinetic theories have the particle relabeling symmetry defining by a right action. This phenomenon can be observed in the following procedure.

The Lie-Poisson structure (\ref{LPBr}) can also be derived by applying the Hamiltonian (Poisson) reduction theorem to the canonical Poisson bracket on the cotangent bundle $T^{\ast }G$ of the Lie group under the lifted action of the group $G$. The cotangent bundle $T^* G$ of a Lie group can be written as the semi-direct product (see e.g. \cite{Fecko})
\begin{equation} \label{tri}
 T^{\ast }G\simeq\mathfrak{g}^*\rtimes G
\end{equation}
of the dual space $\mathfrak{g}^*$ and the group $G$. This is called the right trivialization of the cotangent bundle \cite{KoMiSl93}. In this picture the canonical Poisson bracket (\ref{canPoisson-}) at a point $(\mu,g)$ takes the particular form
\begin{equation}\label{canPoissonG}
  \left\{ F,H\right\}^{T^{\ast}G} = \left\langle T^{\ast}R_{g}\left(F_g\right),H_\mu\right\rangle
-\left\langle T^{\ast}R_{g}\left(H_g\right),F_\mu\right\rangle +\left\langle \mu ,\left[F_\mu,H_\mu \right]\right\rangle,
\end{equation}
see, for example, \cite{Abraham-Marsden}. In this canonical bracket, by taking the functionals $F$ and $H$ free from the group variable $g$, one arrives at the Lie-Poisson bracket (\ref{LPBr}).
The left and right reductions lead to right and left invariant formulations respectively, hence constitutes an interesting geometrical structure called the dual pairs \cite{GaVi12}.

\subsection{Boltzmann equation}
Assume that non-relativistic collision-less plasma rests in a region $Q\subset \mathbb{R}^3$ without boundary (e.g. a torus or vanishing in infinity) with coordinates $\mathbf{r}$, and consider the momentum-phase space $T^\ast Q$ with coordinates $(\mathbf{r,p})$. A one-particle distribution $f$ is a real valued function on the momentum phase space $T^\ast Q$, and the reversible Boltzmann (or Vlasov) equation
\begin{equation}\label{Boltzmann}
\frac{\partial f}{\partial t}+\frac{1}{m} \mathbf{p} \cdot \frac{\partial f}{\partial \rr} -e \frac{\partial \phi}{\partial \rr}\cdot \frac{\partial f}{\partial \pp}=0.
\end{equation}
governs the motion of the plasma. Here, $\phi$ is the electrical potential.

Following \cite{esen2012geometry,Marsden1982}, we now show that how Boltzmann equation (\ref{Boltzmann}) can be written in the form of Lie-Poisson equation (\ref{LP}). The symmetry of the plasma is the relabeling symmetry under the action of canonical transformations $G=Diff_{can}(T^*Q)$. The Lie algebra of this symmetry group can be considered as the space $\mathfrak{g}=\mathcal{F}(T^*Q)$ of smooth functions on $T^\ast Q$ modulo constant functions. Interestingly, the Lie algebra bracket on $\mathfrak{g}=\mathcal{F}(T^*Q)$ is the canonical Poisson bracket (\ref{canPoisson-}). The dual of this space is the space of densities $\mathfrak{g}^*=Den(T^*Q)$ which has elements in form $f(\mathbf{z})d\mathbf{z}$. Here, $f=f(\mathbf{z})$ is a function on and $d\mathbf{z}$ is a top (volume) form on $T^\ast Q$. We fix the symplectic volume $\Omega^3_Q=d \rr d \pp$ as the volume form and define the Lie-Poisson bracket
\begin{equation}\label{eq.PB.B}
\{F,H\}^{(B)}=\int d \rr d \pp f\left(\frac{\partial F_f}{\partial \rr}\cdot\frac{\partial H_f}{\partial \pp} - \frac{\partial H_f}{\partial \rr}\cdot\frac{\partial F_f}{\partial \pp}\right).
\end{equation}
The remaining job is to decide a correct Hamiltonian function generating the reversible Boltzmann (or Vlasov) equation (\ref{Boltzmann}). The Hamiltonian function(al) on $\mathfrak{g}^*$ is
\begin{equation}\label{BoltzmannHam}
H(f)=\int d \rr \diff\pp f(\mathbf{r,p})h(\mathbf{r,p}),
\end{equation}
where $d\mu$ is the top form on $T^\ast Q$ and $h=(1/2m)p^2+ e\phi$ is the particle total energy. Evolution of a functional $F$ on $\mathfrak{g}^*$ is then given by
\begin{eqnarray}
 \dot{F} &=& \{F,H\} = \int d \rr \diff\pp F_f \left(-\frac{\partial f}{\partial \rr} \cdot \frac{\partial h}{\partial \pp} + \frac{\partial f}{\partial \pp}\cdot  \frac{\partial h}{\partial \rr} \right)\nonumber\\
 &=& \int d \rr \diff\pp F_f \left(-\frac{1}{m} \mathbf{p} \cdot \frac{\partial f}{\partial \rr} +e \frac{\partial \phi}{\partial \rr}\cdot \frac{\partial f}{\partial \pp}\right),
\end{eqnarray}
from which equation \eqref{Boltzmann} can be read easily. 

We remark that, we do understood that the potential $\phi$ is externally given. Otherwise, there exists a non-standard fraction $1/2$ in front of the potential $\phi$ if the Poisson equation, which is the gauge invariance of the canonical symplectic formulation on $T^\ast Q$, is coupled to the Vlasov equation \cite{gumral2010geometry,esen2012geometry}. In this case, the coupled system is called Poisson-Vlasov equations. This non-standard fraction $1/2$ is the manifestation of the Green function solution of the Poisson equation \cite{gumral2010geometry}.

\section{Coupling of Two Hamiltonian Systems} \label{CTHS}

In this section, we start with two Lie groups $G$ and $K$ under the mutual interactions. Recall that on the cotangent bundle of a Lie group we have a canonical Poisson bracket \eqref{canPoissonG}, and in the reduced picture (that is on the dual space $\mathfrak{g}^*$) there exists Lie-Poisson bracket \eqref{LPB}. In this section, we present the most general way to couple (match) two canonical Poisson brackets as well as to couple (match) two Lie-Poisson brackets. To achieve these goals, we first exhibit some geometry on the matched pair Lie groups and Lie algebras.

\subsection{Matched Pair of Lie Groups and Lie algebras}

Let $G$ be a Lie group with identity element $e$ and $\mathcal{M}$ be a set. The left group action of $G$ on $\mathcal{M}$ is a differentiable mapping  $$ G\times \mathcal{M} \mapsto \mathcal{M}: (g,x)\mapsto g\vartriangleright x$$
satisfying the identity condition $e\vartriangleright x=x$ for all $x$ in $\mathcal{M}$ and the associativity condition $$g\vartriangleright (h\vartriangleright x)=(gh) \vartriangleright x$$ for all $g$, $h$ in $G$ and all $x$ in $\mathcal{M}$, see, for example, \cite{Fecko}.

Let $(G,K)$ be a pair of Lie groups, such that $K$ acts on $G$ from the left,
and $G$ acts on $K$ from the right by
\begin{subequations} \label{rho}
\begin{eqnarray}
\rho &:&K\times G\rightarrow G:\left( h,g\right) \mapsto
h\vartriangleright g, \\ \sigma &:&H\times G\rightarrow K: \left(
h,g\right) \mapsto h\vartriangleleft g.
\end{eqnarray}
\end{subequations}
The pair $(G,K)$ is called a matched pair of Lie groups if the mutual actions
\eqref{rho} satisfy the compatibility conditions
\begin{subequations} \label{comp}
\begin{eqnarray}
h\vartriangleright \left( g_{1}g_{2}\right) &=&\left( h\vartriangleright
g_{1}\right) \left( \left( h\vartriangleleft g_{1}\right) \vartriangleright
g_{2}\right) \mbox{ and} \\
(h_{1}h_{2})\vartriangleleft g&=&\left(
h_{1}\vartriangleleft \left( h_{2}\vartriangleright g\right) \right) \left(
h_{2}\vartriangleleft g\right)
\end{eqnarray}
\end{subequations}
along with $h\vartriangleright e_G = e_G$ and $e_K\vartriangleleft g = e_K$. Here, $e_G \in G$ and $e_K\in K$ are the identity elements. In this case, the Cartesian product $G\times K$ becomes a Lie group with the multiplication
\begin{equation} \label{mp}
\left( g_{1},h_{1}\right) \left(
g_{2},h_{2}\right) =\left( g_{1}\left( h_{1}\vartriangleright
g_{2}\right) ,\left( h_{1}\vartriangleleft g_{2}\right) h_{2}\right),
\end{equation}
and is denoted by $G\bowtie K$. A matched pair Lie group $G\bowtie K$ is a Lie group containing $G$ and $K$ as two non-intersecting Lie subgroups in mutual actions satisfying the compatibility conditions (\ref{comp}), see \cite{LuWein90,Maji90-II,Maji90,Majid-book,Take81}. The operation $\bowtie$ is called the matched pair or the bicross product in the literature. We point out the presence of mutual actions in the group multiplication (\ref{mp}) are crucial while coupling two interacting Hamiltonian systems.

If one of the actions in \eqref{rho} is trivial, then the matched pair group
$G\bowtie K$ reduces to a semi-direct product group. Explicitly, if only $G$ acts on $K$, then the group is called a semi-direct product Lie group $G\ltimes K$ and the group operation given in (\ref{mp}) reduces to
\begin{equation} \label{sd1}
\left( g_{1},h_{1}\right) \left(
g_{2},h_{2}\right) =\left( g_{1} g_{2},\left( h_{1}\vartriangleleft g_{2}\right) h_{2}\right).
\end{equation}
If only $K$ acts on $G$, then the group is the semi-direct product Lie group $G\rtimes K$ and the group operation given in (\ref{mp}) reduces to
\begin{equation} \label{sd2}
\left( g_{1},h_{1}\right) \left(
g_{2},h_{2}\right) =\left( g_{1}\left( h_{1}\vartriangleright
g_{2}\right) ,h_{1} h_{2}\right).
\end{equation}
When both of the actions are trivial, we arrive at the direct product Lie group $G\times K$ with the group multiplication
\begin{equation} \label{dp}
\left( g_{1},h_{1}\right) \left(
g_{2},h_{2}\right) =\left( g_{1} g_{2},h_{1} h_{2}\right).
\end{equation}

Let $\mathfrak{g}$ and $\mathfrak{k}$ be two Lie algebras in mutual interaction over a common base field. We write the actions as 
\begin{subequations}\label{Lieact}
\begin{eqnarray}
\vartriangleright&:&\mathfrak{k}\times\mathfrak{g}\rightarrow \mathfrak{g}:(\eta,\xi) \mapsto \eta\vartriangleright \xi, \\ \vartriangleleft&:&\mathfrak{k}\times\mathfrak{g}\rightarrow \mathfrak{k}: (\eta,\xi) \mapsto \eta\vartriangleleft \xi.
\end{eqnarray}
\end{subequations}
The pair $(\mathfrak{g},\mathfrak{k})$ is called a matched pair of Lie algebras if the compatibilities
\begin{eqnarray}\label{compinf}
\eta \vartriangleright \lbrack \xi _{1},\xi _{2}]&=&[\eta \vartriangleright
\xi _{1},\xi _{2}]+[\xi _{1},\eta \vartriangleright \xi _{2}]+(\eta
\vartriangleleft \xi _{1})\vartriangleright \xi _{2}-(\eta \vartriangleleft
\xi _{2})\vartriangleright \xi _{1}, \\
\lbrack \eta _{1},\eta _{2}]\vartriangleleft\xi &=&[\eta _{1},\eta _{2}\vartriangleleft\xi ]+[\eta _{1}%
\vartriangleleft\xi ,\eta _{2}]+\eta _{1}\vartriangleleft (\eta _{2}\vartriangleright \xi
)-\eta _{2}\vartriangleleft (\eta _{1}\vartriangleright \xi )
\end{eqnarray}
are satisfied. In this case, the sum $\mathfrak{g} \oplus \mathfrak{k}$ is a Lie algebra with the Lie bracket
\begin{equation}
\lbrack (\xi _{1},\eta _{1}),\,(\xi _{2},\eta _{2})]=\left( [\xi _{1},\xi
_{2}]+\eta _{1}\vartriangleright \xi _{2}-\eta _{2}\vartriangleright \xi
_{1},\,[\eta _{1},\eta _{2}]+\eta _{1}\vartriangleleft \xi _{2}-\eta
_{2}\vartriangleleft \xi _{1}\right),  \label{mpla}
\end{equation}%
and is denoted by  $\mathfrak{g\bowtie k}$. It is immediate to see that, in this case, the constitutive Lie algebras $\mathfrak{g}$ and $\mathfrak{k}$ are trivially intersecting Lie subalgebras of the matched pair. We remark once more that, the existence of the mutual infinitesimal actions (\ref{Lieact}) in the Lie algebra bracket (\ref
{mpla}) will play a prominent role while coupling two Hamiltonian systems.

The compatibility conditions in (\ref{compinf}) are the infinitesimal versions of the group compatibility conditions (\ref{comp}). If the matched pair Lie algebra is defined without referring to a matched pair Lie group then they should be checked. If the matched pair Lie algebra is derived from a matched pair Lie group then the conditions (\ref{compinf}) are automatically satisfied. In this case,  $\mathfrak{g}$ is the Lie algebra of the group $G$, and $\mathfrak{k}$ is the Lie algebra of $K$. The infinitesimal actions in (\ref{Lieact}) are obtained by deriving the group actions in (\ref{rho}).

If one of the actions in (\ref{Lieact}) are trivial, one arrives at the semidirect product Lie algebra structure. Namely, if the action of $\mathfrak{g}$ on $\mathfrak{k}$ is trivial then the matched pair Lie algebra (\ref{mpla}) reduces to the semidirect product Lie algebra multiplication
\begin{eqnarray}
\lbrack (\xi _{1},\eta _{1}),\,(\xi _{2},\eta _{2})]=\left( [\xi _{1},\xi
_{2}]+\eta _{1}\vartriangleright \xi _{2}-\eta _{2}\vartriangleright \xi
_{1},\,[\eta _{1},\eta _{2}]\right)
\end{eqnarray}
on $\mathfrak{g\rtimes k}$, whereas if the action of $\mathfrak{k}$ on $\mathfrak{g}$ is trivial then the matched pair Lie algebra structure (\ref{mpla}) reduces to
\begin{equation}
\lbrack (\xi _{1},\eta _{1}),\,(\xi _{2},\eta _{2})]=\left( [\xi _{1},\xi
_{2}],[\eta _{1},\eta _{2}]+\eta _{1}\vartriangleleft \xi _{2}-\eta
_{2}\vartriangleleft \xi _{1}\right),  \label{mpla2}
\end{equation}%
on $\mathfrak{g\ltimes k}$. If both of the actions are trivial then one arrives the direct product Lie algebra structure
\begin{eqnarray}
\lbrack (\xi _{1},\eta _{1}),\,(\xi _{2},\eta _{2})]=\left( [\xi _{1},\xi
_{2}],[\eta _{1},\eta _{2}]\right)
\end{eqnarray}
on $\mathfrak{g\times k}$.

The dual of the matched pair Lie algebra $\mathfrak{g\bowtie k}$ is given by the Cartesian product $\mathfrak{g^*\times k^*}$ of the dual spaces $\mathfrak{g^*}$ and $\mathfrak{k^*}$. The dualization is given by
$$ (\mathfrak{g^*\times k^*})\times (\mathfrak{g\bowtie k}):((\mu,\nu),(\xi,\eta))\rightarrow
\langle \mu,\xi \rangle + \langle \nu, \eta \rangle. $$
Recall that we have presented the cotangent bundle of a Lie group by the product of the group and the dual space in (\ref{tri}). By following the same understanding, we now identify the cotangent bundle $T^{\ast}(G\bowtie
K)$ with its right trivialization
\begin{equation} \label{tri2}
  T^{\ast}(G\bowtie
H)\simeq (\mathfrak{g}^\ast \times \mathfrak{k}^\ast)\rtimes (G\bowtie K).
\end{equation}
Under this identification, an element of $T^{\ast}(G\bowtie
K)$ can be represented by a four-tuple $(\mu,\nu,g,h) \in (\mathfrak{g}^\ast \times \mathfrak{k}^\ast)\rtimes (G\bowtie K)$.

\subsection{Matched Pair of Poisson Brackets}

Being a cotangent bundle, $T^{\ast}(G\bowtie
K)$ is a symplectic manifold hence it is equipped with the canonical Poisson bracket $\left\{ \bullet,\bullet\right\}^{T^{\ast}(G\bowtie K)}$. Although it has technically the same structure with the canonical Poisson bracket $\left\{ \bullet,\bullet\right\}^{T^{\ast}G}$ on $T^{\ast}G$ presented in (\ref{canPoissonG}), this time the trivialization (\ref{tri}) enables us to recast the canonical Poisson bracket on $T^{\ast}(G\bowtie
K)$ in terms of actions and the canonical Poisson structures on $T^{\ast}G$ and $T^{\ast}K$. 

At the point $(\mu,\nu,g,h)$, the canonical Poisson bracket is given by
\begin{eqnarray}  \label{PoissononT*GH2}
\left\{ F,H\right\}^{T^{\ast}(G\bowtie
K)} =\underbrace{\left\langle T^{\ast}R_{g}\left(F_g\right),H_\mu\right\rangle
-\left\langle T^{\ast}R_{g}\left(H_g\right),F_\mu\right\rangle +\left\langle \mu ,\left[F_\mu,H_\mu \right]\right\rangle}_{\text{ Canonical Poisson bracket on $\mathfrak{g}^{\ast}\rtimes G$}%
} \notag \\
\underbrace{+\left\langle F_g,H_\nu \vartriangleright g \right\rangle -\left\langle H_g,F_\nu \vartriangleright g \right\rangle +\left\langle \nu ,\left[F_\nu,H_\nu
\right]\right\rangle}_{\text{ Poisson bracket on $\mathfrak{k}^{\ast}\rtimes G$}}  \notag
\\
\underbrace{ +\left\langle
T^{\ast}R_h\left(F_h\right) ,H_\nu\vartriangleleft g \right\rangle -\left\langle
T^{\ast}R_h\left(H_h\right) ,F_\nu\vartriangleleft g \right\rangle}_{\text{ Terms by
the action of $G$ on $T^{\ast}H$}} \notag
\\+
\underbrace{ \left\langle \nu, H_\nu \vartriangleleft F_\mu -  F_\nu \vartriangleleft H_\mu \right\rangle }_{\text{ Terms by the action
of $\mathfrak{g}$ on $\mathfrak{k}^\ast$}}
\underbrace{+ \left\langle \mu , F_\nu\vartriangleright H_\mu - H_\nu \vartriangleright F_\mu\right\rangle }_{\text{ Terms by
the action of $\mathfrak{k}$ on $\mathfrak{g}^\ast$}},
\end{eqnarray}
where $F_g$ is an element of $T^\ast _g G$, $T^\ast R$ is the cotangent lift of the right translation map on the group level, $H_\nu \vartriangleright g$ is the infinitesimal action of $\mathfrak{k}$ on $G$, $F_\nu \vartriangleleft g$ is the lift of the action of $G$ on $\mathfrak{k}$, and $F_\mu$ is an element of $\mathfrak{g}$, \cite{esen2016hamiltonian}. Note that while presenting the matched canonical Poisson bracket (\ref{PoissononT*GH2}), we have combined some of the terms and labeled them. The first line is the canonical Poisson bracket on $T^\ast _g G$ also given in (\ref{canPoissonG}). If the function(al)s depend only on $g$ and $\nu$ in the expression of the matched canonical Poisson bracket (\ref{PoissononT*GH2}), remains the second line remains, which is the Poisson bracket on $\mathfrak{k}^{\ast}\rtimes G$. As we shall point out later on, existence of this reduced Poisson bracket enables one to write magnetohydrodynamics and electrohydrodynamics as two subsystems of the electromagnetohydrodynamics in a geometric way.

Perhaps the most important particular case of the matched canonical Poisson bracket is the matched Lie-Poisson bracket on the dual space $\mathfrak{g}^\ast\times\mathfrak{k}^\ast$, given by
\begin{align}  \label{LiePoissonongh}
& \left\{F,H\right\} _{\mathfrak{g}^\ast\times\mathfrak{k}%
^\ast}(\mu,\nu) =\underbrace{\left\langle \mu ,\left[F_\mu,H_\mu\right]\right\rangle
+\left\langle \nu ,\left[F_\nu,H_\nu\right]\right\rangle}_{\text{Direct product}} +\\
& \hspace{2cm} \underbrace{\left\langle \mu ,F_\nu%
\vartriangleright H_\mu\right\rangle
-\left\langle \mu ,H_\nu\vartriangleright
F_\mu\right\rangle}_ {\text{Due to the action of $\mathfrak{k}$ on $\mathfrak{g}$}} +\underbrace{ \left\langle \nu ,F_\nu%
\vartriangleleft H_\mu\right\rangle
-\left\langle \nu ,H_\nu\vartriangleleft
F_\mu\right\rangle} _{\text{Due to the action of $\mathfrak{g}$ on $\mathfrak{k}$}}.  \notag
\end{align}
The first term is naive coupling of two Lie-Poisson brackets in form (\ref{LPBr}). If there is no interaction, that is if the group actions (\ref{rho}) are trivial or the infinitesimal actions (\ref{Lieact}) are zero action, then the Lie-Poisson bracket for the coupled system is given only by the first direct product term. If there exists only one sided action, say the action of the Lie algebra $\mathfrak{k}$ on $\mathfrak{g}$ that is semidirect product $\mathfrak{g\rtimes k}$, then the first two terms determine the Poisson bracket. This shows how the matched pair dynamics covers the semi direct product theory.

As pointed out previously, there exist two important implications of the matched Poisson brackets (\ref{PoissononT*GH2}) and (\ref{LiePoissonongh}). Firstly, they show how to couple two systems in mutual actions. Secondly, if the configuration space can be written as the matched pair of its two subgroups, they show how to decouple a system in to its two of its subsystems in a purely geometrical way. We refer to an upcoming study \cite{esen2016vlasov} for the matched pair decomposition of the Boltzmann bracket (\ref{Boltzmann}).

\subsection{Semidirect Product Theory}
Let $V$ be a vector space. The tangent and the cotangent bundles are given by $TV =V \times V$ and $T^\ast V =V \times V^\ast $. We consider a left action of a Lie group $K$ on the vector space $V$ denoted by
\begin{equation} \label{sdv}
 \varphi: K\times V \to V: (h,\mathbf{v}) \mapsto h\vartriangleright\mathbf{v}.
\end{equation}
In this case, the product of $K$ and $V$ is a Lie group called the semidirect product group and denoted by $K\ltimes V$ \cite{marsden1998symplectic,marsden1984semidirect,marsden1984reduction,ratiu1980motion}. This is the particular case of the semidirect product group presented in (\ref{sd2}), where  $G$ be the vector space $V$ hence $\mathfrak{g}=V$. We denote the corresponding infinitesimal action by
\begin{equation}
\mathfrak{k}\times V \mapsto V: (\eta, \mathbf{v}) \mapsto \eta\vartriangleright \mathbf{v}.
\end{equation}
Let us now recall the matched canonical Poisson bracket (\ref{PoissononT*GH2}) and to adapt this particular case to it. Under the infinitesimal action of the Lie algebra $\mathfrak{k}$ on $T^\ast V$, the coupling $\left\{ \bullet,\bullet\right\}^{\mathfrak{k}^\ast \times T^\ast V}$ of the canonical Poisson bracket $\left\{ \bullet, \bullet \right\}^{T^\ast V}$ on $T^\ast V$ and the Lie Poisson bracket $\left\{ \bullet, \bullet \right\}^{\mathfrak{k}^*}$ on $\mathfrak{k}^\ast$ (for two functionals $F=F(\nu,\mathbf{v},\mathbf{\alpha})$ and $H=H(\nu,\mathbf{v},\mathbf{\alpha})$) can be computed as
\begin{align}  \label{PoissononT*GH}
&\left\{ F,H\right\}^{\mathfrak{k}^\ast \times T^\ast V}= \left\{ F,H\right\}^{T^\ast V} + \left\{ F,H\right\}^{\mathfrak{k}^\ast} \\&
+\left\langle  F_\mathbf{v},H_\nu\vartriangleright\mathbf{v}\right\rangle
-\left\langle H_\mathbf{v},F_\nu\vartriangleright\mathbf{v}\right\rangle
+ \left\langle \mathbf{\alpha} ,F_\nu \vartriangleright H_ \mathbf{\alpha}\right\rangle
-\left\langle \mathbf{\alpha} ,H_\nu\vartriangleright
F_\mathbf{\alpha}\right\rangle. \notag
\end{align}
Note that, the additional terms in the second line are manifesting the action of $K$ on the vector space $V$ and its dual $V^*$.

In particular, if the functionals are independent of $\mathbf{v}$, this bracket reduces to the Lie-Poisson bracket on  $\mathfrak{k}^\ast\times V^\ast $ given by
\begin{eqnarray} \label{LP.br}
\left\{F,H\right\}^{\mathfrak{k}^\ast\times V^\ast } &=& \{F,H\}^{\mathfrak{k}^\ast} +\left\langle \alpha ,F_\nu
\vartriangleright H_\alpha\right\rangle
-\left\langle \alpha ,H_\nu\vartriangleright
F_\alpha\right\rangle .  \notag
\end{eqnarray}
On the other hand, the projection of the bracket (\ref{PoissononT*GH}) onto the product space $\mathfrak{k}^\ast \times V$ is a Poisson bracket, for two function(al)s $F=F(\nu,\mathbf{v})$ and $H=H(\nu,\mathbf{v})$  given by
\begin{align}  \label{eq.PBh*V}
&\left\{ F,H\right\}^{\mathfrak{k}^\ast \times V}= \left\{ F,H\right\}^{\mathfrak{k}^\ast} +\left\langle  F_\mathbf{v},H_\nu\vartriangleright\mathbf{v}\right\rangle
-\left\langle H_\mathbf{v},F_\nu\vartriangleright\mathbf{v}\right\rangle.
\end{align}
We remark that this bracket is not in a Lie-Poisson form. Instead, it can be non-degenerate if the action (\ref{sdv}) is free and transitive.

\section{The Hydrodynamics Equations}

\subsection{Classical hydrodynamics} \label{sec.HD}
Assume that a continuum is present in $\mathcal{Q}\subset \mathbb{R}^3$ without boundary (or where all variables vanish near infinity). The configuration space for the continuum can be considered as the group $Diff(\mathcal{Q})$ of diffeomorphisms on $\mathcal{Q}$. This is an infinite dimensional Lie group. Its Lie algebra is the space $\mathfrak{X}(\mathcal{Q})$ of smooth vector fields on $\mathcal{Q}$.

In order to define the Lie algebra bracket on $\mathfrak{X}(\mathcal{Q})$, let us recall the following. The left infinitesimal action of a vector field $X\in \mathfrak{X}(\mathcal{Q})$ on a function $\sigma \in \mathcal{F}(\mathcal{Q})$ is defined simply by taking the directional derivative of $\sigma$ in the direction of $X$. In a local coordinate system $\mathbf{r}=(r^i)$, a vector field is in the form $X=X^i(r) \partial _i$, and then
\begin{equation} \label{act}
  \mathfrak{X}(\mathcal{Q})\times\mathcal{F}(\mathcal{Q})\to\mathcal{F}(\mathcal{Q}):(X,\sigma)\mapsto -X(\sigma)=-X^i \sigma_i,
\end{equation}
where we assume the summation on repeated indices and the abbreviated notation $\sigma_i:=\partial \sigma / \partial r^i$ for the partial derivatives. Using this action, one defines the Jacobi-Lie bracket of vector fields $X$ and $Y$ given by
\begin{equation} \label{Lie1}
  \mathcal{L}_X (Y) (\sigma)= [X,Y](\sigma) = X(Y(\sigma)) - Y(X(\sigma)).
\end{equation}
The Lie algebra bracket on $\mathfrak{X}(\mathcal{Q})$ is minus the Jacobi Lie bracket of vector fields.

The dual space of $\mathfrak{X}(\mathcal{Q})$ is the space of one-form densities $\Lambda^1(\mathcal{Q})\times Den(\mathcal{Q})$and the pairing is given by simply the multiply-and-integrate formula. For a vector field $X=X^i(r)\partial _i$ and a one-form density $u=u_i(r)dr^i$, the pairing is given by
\begin{equation} 
  \int_{\mathcal{Q}}d\mathbf{r} u(r) \cdot X (r).
\end{equation}
Some functional analytic issues should be considered for the convergence of such integrals. Instead, we are assuming that proper functional spaces are chosen in order to guarantee the existence of the integrals \cite{arnold1999topological,ebinmarsden}.

To write the equation of the motion of the hydrodynamics, we first define the semidirect product space $\mathfrak{X}(\mathcal{Q})\ltimes(\mathcal{F}(\mathcal{Q}) \times \mathcal{F}(\mathcal{Q}))$ consisting of the space $\mathfrak{X}(\mathcal{Q})$ of vector fields and two copies of the space $\mathcal{F}(Q)$ of smooth functions on $\mathcal{Q}$. An action of a vector field on the smooth function is defined as in (\ref{act}). Accordingly, the semi-direct product Lie algebra structure on $\mathfrak{X}(\mathcal{Q})\ltimes(\mathcal{F}(\mathcal{Q}) \times \mathcal{F}(\mathcal{Q}))$  is given by
$$[(X,\sigma_1,\sigma_2),(Y,\beta_1,\beta_2)]=(-[X,Y],Y( \sigma_1) - X(\beta_1), Y( \sigma_2)-X( \beta_2)).$$
We fix a volume (top) form $d\mathbf{r}$ on $\mathcal{Q}$, so that we can take the dual space as $\Lambda^1(\mathcal{Q})\times(\mathcal{F}(\mathcal{Q})\times \mathcal{F}(\mathcal{Q}))$, where $\Lambda^1(\mathcal{Q})$ is the space of one-forms. Note that, in this definition we have identified the dual $\mathcal{F}^*(\mathcal{Q})$ of $\mathcal{F}(\mathcal{Q})$ by considering the integration ($L^2$-pairing) as a weakly non-degenerate inner product on $\mathcal{F}(\mathcal{Q})$. As a result, an element of the dual space is a three tuple $(M,\rho,s)$, where $M$ is the momenta, $\rho$ is the mass density, and $s$ is the entropy. A direct calculation\footnote{using that $F_u$ is a vector field, that acts on scalars as $F_{u_i}\partial_i H_\rho$,} shows that, the Lie-Poisson bracket (\ref{LP.br}) takes the particular form \cite{marsden1984semidirect},
\begin{eqnarray}\label{eq.PB.CH}
 &&\{F,H\}^{(CH)}(u,\rho,s)=\int d\rr \rho \left((H_u\cdot \nabla) F_\rho - (F_u\cdot \nabla) H_\rho\right)\nonumber\\
 &&+ \int d\rr u \cdot ((H_u\cdot \nabla) F_u- (F_u\cdot \nabla) H_u) \nonumber \\ &&+ \int d\rr s \left((H_u\cdot \nabla) F_s - (F_u\cdot \nabla) H_s\right).
\end{eqnarray}
In order to generate the Hamilton's equations governing the hydrodynamics, one introduces the Hamiltonian function(al) as the total energy given by
\begin{equation}
 H(u,\rho,s) = \int d\rr \frac{u^2}{2\rho}+ \eps(\rho,s).
\end{equation}
In this case, the Hamilton's equations turn out to be
\begin{subequations}
\begin{eqnarray} {\label{eqn.PB.CH}}
 \frac{\partial \rho}{\partial t} &=& -\partial_i u_i, \\
 \frac{\partial u_i}{\partial t}  &=& -\partial_j\left(\frac{u_i u_j}{\rho}\right) - \rho \partial_i \eps_{\rho} - s \partial_i \eps_{s}, \\
 \frac{\partial s}{\partial t} &=& -\partial_i \left(s \frac{u_i}{\rho}\right),
\end{eqnarray}
\end{subequations}
which represent the Euler equations for compressible ideal fluid.

It is also possible to arrive at the level of hydrodynamics from the level of Boltzmann easily as follows. The projections
\begin{subequations}
\begin{eqnarray}
 \rho(\rr_a)&=&\int d\rr \int d\pp m f(\rr,\pp)\delta(\rr-\rr_a) \label{PB.eq.CH.rho}\\
 u_i(\rr_a)&=&\int d\rr \int d\pp p_i f(\rr,\pp)\delta(\rr-\rr_a)\label{PB.eq.CH.u}\\
 \label{PB.eq.CH.s}s(\rr_a)&=&\int d\rr \int d\pp \sigma(f(\rr,\pp))\delta(\rr-\rr_a)
\end{eqnarray}
\end{subequations}
are called plasma-to-fluid mappings from the Boltzmann Poisson bracket \eqref{eq.PB.B} to the hydrodynamics Poisson bracket (\ref{eq.PB.CH}), \cite{Marsden-Ratiu,MaWeRaScSp83,PhysicaD-2015}. Here, $m$ is mass of one particle and entropy density $\sigma$ is a positive smooth real-valued function of the distribution function. In this level, the relationship between velocity $\mathbf{v}$ and the momentum $M$ is simply $u=\rho\mathbf{v}$.

\subsection{Binary hydrodynamics}
Consider a mixture of two fluids described by state variables $(u^1, u^2,\rho_1, \rho_2, s_1, s_2)$, see e.g. \cite{PhysicaD-2015}. That means that the mixture is described by density, momentum density and entropy density of each constituent. The reversible evolution of these variables is generated by direct product Poisson bracket
\begin{eqnarray}\label{eq.PB.BHD}
 &&\{F,H\}^{(BH)}=\{F,H\}^{(CH)_1} + \{F,H\}^{(CH)_2} \\
 &&=\int d\rr  \rho_1 \left((H_{u_1}\cdot \nabla) F_{\rho_1} - (F_{u_1}\cdot \nabla) H_{\rho_1}\right)\nonumber \\
 &&+ \int d\rr  \rho_2 \left((H_{u_2}\cdot \nabla) F_{\rho_2} - (F_{u_2}\cdot \nabla) H_{\rho_2}\right)\nonumber \\
 &&+ \int d\rr u_1\cdot ((H_{u_1}\cdot \nabla) F_{u_1}- (F_{u_1}\cdot \nabla) H_{u_1})\nonumber \\
 &&+
 \int d\rr u_2\cdot ((H_{u_2}\cdot \nabla) F_{u_2}- (F_{u_2}\cdot \nabla) H_{u_2}) \nonumber
 \\ && + \int d\rr s_1 \left((H_{u_1}\cdot \nabla) F_{s_1} - (F_{u_1}\cdot \nabla) H_{s_1}\right)
 \nonumber \\
 &&+ \int d\rr s_2 \left((H_{M_2}\cdot \nabla) F_{s_2} - (F_{M_2}\cdot \nabla) H_{s_2}\right) .
\end{eqnarray}
This bracket can be also obtained by simple projection from binary Boltzmann Poisson bracket, which can be obtained by projection from Liouville Poisson bracket, see \cite{PhysicaD-2015}.
We take the Hamiltonian as the energy function
\begin{equation}
 E = \int d \rr \frac{(u^1)^2}{2\rho_1}+ \frac{(u^2)^2}{2\rho_2} + \eps(\rho_1, \rho_2, s_1, s_2).
\end{equation}
Then the Hamilton's equations for the binary hydrodynamics are computed to be
\begin{subequations}
\begin{eqnarray}
 \frac{\partial \rho_1}{\partial t} &=& -\partial_i u^1_i\\
 \frac{\partial \rho_2}{\partial t} &=& -\partial_i u^2_i\\
 \frac{\partial u^1_i}{\partial t} &=& -\partial_j\left(\frac{u^1_i u^1_j}{\rho_1}\right) - \rho_1 \partial_i \eps_{\rho_1} - s_1 \partial_i \eps_{s_1}\\
 \frac{\partial u^2_i}{\partial t} &=& -\partial_j\left(\frac{u^2_i u^2_j}{\rho_2}\right) - \rho_2 \partial_i \eps_{\rho_2} - s_2 \partial_i \eps_{s_2}\\
 \frac{\partial s_1}{\partial t} &=& -\partial_i \left(s_1 \frac{u^1_i}{\rho_1}\right)\\
 \frac{\partial s_2}{\partial t} &=& -\partial_i \left(s_1 \frac{u^2_i}{\rho_2}\right)
\end{eqnarray}
\end{subequations}

Let us define total momentum and total entropy of the binary system as
\begin{subequations}
\begin{eqnarray}
 u &=& u^1 + u^2,\\
 s &=& s_1 + s_2,
\end{eqnarray}
\end{subequations}
respectively. This enables us to perform a Hamiltonian reduction procedure from two copies of $\Lambda^1(\mathcal{Q})\times(\mathcal{F}(\mathcal{Q})\times \mathcal{F}(\mathcal{Q}))$ to the product space $\Lambda^1(\mathcal{Q})\times(\mathcal{F}(\mathcal{Q})\times (\mathcal{F}(\mathcal{Q}))) \times \mathcal{F}(\mathcal{Q})$ with coordinates $(u,\rho_1, \rho_2, s)$. Hence, after a direct calculation the direct product Poisson bracket \eqref{eq.PB.BHD} takes the following reduced form
\begin{eqnarray}
&&\{F,H\}^{(CBH)}=\int d\rr  \rho_1 \left((H_{u}\cdot \nabla) F_{\rho_1} - (F_{u}\cdot \nabla) H_{\rho_1}\right)\nonumber \\ 
&&+ \int d\rr  \rho_2 \left((H_{u}\cdot \nabla) F_{\rho_2} - (F_{u}\cdot \nabla) H_{\rho_2}\right)\nonumber \\
 &&+ \int d\rr u \cdot ((H_u\cdot \nabla) F_u- (F_u\cdot \nabla) H_u) \nonumber \\ &&+ \int d\rr s \left((H_u\cdot \nabla) F_s - (F_u\cdot \nabla) H_s\right)
\end{eqnarray}
which we call classical binary hydrodynamics bracket expressing reversible motion of binary mixtures within Classical Irreversible Thermodynamics (CIT) \cite{dGM}. A further reduction to $(u, \rho = \rho_1+\rho_2,s)$ leads to the hydrodynamic Poisson bracket \eqref{eq.PB.CH}.

\section{Electromagnetic Field and Matter}

In this section, we shall study the theoretical results derived in the previous section in the particular case of the various couplings of motion of matter electromagnetic field.

\subsection{Kinetic electrodynamics}\label{sec.KED}
To couple the Maxwell equations with the reversible Boltzmann equation, we start by taking the direct product of Poisson brackets (\ref{eq.PB.ED.Can}) and (\ref{eq.PB.B}). To formulate the resulting Poisson bracket in terms of $f$, $\mathbf{E}$ and $\mathbf{B}$ instead of the canonical electromagnetic variables, $\mathbf{A}$ and $\mathbf{Y}$, one needs to employ gauge invariance in the velocity formulation. As a result \cite{Marsden1982, Marsden-Ratiu}, one has
\begin{eqnarray}\label{eq.PB.KED}
 \{F,G\}^{(KED)}&=&\{F,G\}^{(B)} + \{F,G\}^{(EM)}+\nonumber\\
 &&+\int d\rr \int d\pp \frac{ze}{\eps_0} \frac{\partial f}{\partial \pp}\cdot\left(F_{\EE} G_f - G_{\EE}F_f\right)\nonumber\\
 &&+\int d\rr \int d\pp z e f \mathbf{B}\cdot\left(\frac{\partial F_f}{\partial \pp} \times \frac{\partial G_f}{\partial \pp}\right).
\end{eqnarray}
 Charge number per particle and elementary charge are denoted by $z$ and $e$, respectively. The (KED)-bracket \eqref{eq.PB.KED} is also endowed with the constraints
\begin{subequations}
\begin{eqnarray}
 \dive \EE &=& \frac{z e}{\eps_0}\intd\pp f, \\
 \dive \BB &=& 0.
\end{eqnarray}
\end{subequations}
See \cite{cendra1998maxwell} for the Lagrangian formulation of this system.

\subsection{Binary kinetic electrodynamics}\label{sec.BKED}
A Poisson bracket for binary kinetic electrodynamics in terms of by state variables $(f_1, f_2, \EE, \BB)$ is given by
\begin{eqnarray}\label{eq.PB.BKED}
 \{F,H\}^{(BKED)}&=&\{F,H\}^{(B)_1} + \{F,H\}^{(B)_2}+ \{F,H\}^{(EM)}+\nonumber\\
 &&+\sum_{\alpha=1}^2\int d\rr \int d\pp \frac{z_\alpha e}{\eps_0} \frac{\partial f_\alpha}{\partial \pp}\cdot\left(F_{\EE} H_{f_\alpha} - H_{\EE}F_{f_\alpha}\right)\nonumber\\
 &&+\sum_{\alpha = 1}^2 \int d\rr \int d\pp z_\alpha e f_\alpha B\cdot\left(\frac{\partial F_{f_\alpha}}{\partial \pp} \times \frac{\partial H_{f_\alpha}}{\partial \pp}\right)
\end{eqnarray}
where $z_\alpha$ are the respective charge numbers per particle.
\begin{subequations}
This bracket is also endowed with the following constraints
\begin{eqnarray}
 \dive \EE &=& \sum_{\alpha = 1}^2 \frac{z_\alpha e}{\eps_0}\intd\pp f_\alpha, \\
 \dive \BB &=& 0.
\end{eqnarray}
\end{subequations}
Note that while arriving at the binary version, we have simply doubled the terms in (KED)-bracket while fixing (EM)-bracket.

\subsection{Electromagnetohydrodynamics}\label{sec.EMHD}

We will couple the hydrodynamics with the electromagnetic field. Recall from the section {\ref{sec.ED}} that the Maxwell equations are in the canonical form on the cotangent bundle $T^\ast\mathfrak{U}$, and also recall from section \ref{sec.HD} that the hydrodynamics equations are in the Lie-Poisson form on the dual of $\mathfrak{X}(\mathcal{Q})\ltimes(\mathcal{F}(\mathcal{Q}) \times \mathcal{F}(\mathcal{Q}))$. Note that the momentum density will be denoted by $M$, since the physical meaning will be the total momentum
\begin{equation}
M = u + \eps_0 \mathbf{E} \times \mathbf{B},
\end{equation}
which is the sum of hydrodynamic and electromagnetic momentum, see \cite{Landau2} for the electromagnetic momentum. The hydrodynamic Poisson bracket will be the bracket \eqref{eq.PB.CH} with $u$ replaced by $M$.

The matched (coupled) Poisson bracket for the canonical system on $T^\ast V$ and the Lie-Poisson system on $\mathfrak{k}^\ast$ has been exhibited in \eqref{PoissononT*GH}. In the present abstract setting, we particularly take $\mathfrak{k}=\mathfrak{X}(\mathcal{Q})\ltimes(\mathcal{F}(\mathcal{Q}) \times \mathcal{F}(\mathcal{Q}))$ and the vector space as $V=\mathfrak{U}$. The infinitesimal action of $\mathfrak{k}$ on $V$ turns out to be
 \begin{equation*}
\varphi:(\mathfrak{k},V)\to V: ((X,a,b),\mathbf{A})\mapsto X\vartriangleright\mathbf{A}=(- \mathcal{L}_X \mathbf{A}),
 \end{equation*}
 where $\mathcal{L}$ is the Lie derivative of the one-form $\mathbf{A}$. In a local coordinate system, we take $\mathbf{A}=\mathbf{A}_i dr^i$ and $X=X^i \partial_i$, minus the Lie derivative computed to be
\begin{equation} \label{Lie2}
X\vartriangleright\mathbf{A}= - \mathcal{L}_X \mathbf{A}=-(X^j\partial_j \mathbf{A}_i+\mathbf{A}_j\partial_i X^j)dx^i.
 \end{equation}

On the dual space  $\mathfrak{k}^\ast\times T^\ast \mathfrak{U}$, there exists a Poisson bracket given implicitly in (\ref{PoissononT*GH}). To find its explicit version, we perform the following calculation
\begin{eqnarray} \label{EMHDc}
\left\{F,H\right\}^{(EMHDc)}&=& \{F,H\}^{(CH)} + \{F,H\}^{(EMc)}+\left\langle  F_\mathbf{A},H_{(M,\rho,s)}\vartriangleright \mathbf{A}\right\rangle \notag \\ &&
-\left\langle H_\mathbf{A},F_{(M,\rho,s)}\vartriangleright \mathbf{A}\right\rangle
+ \left\langle \mathbf{\mathbf{Y}} ,F_{(M,\rho,s)} \vartriangleright H_ \mathbf{\mathbf{Y}}\right\rangle
\notag  \\ &&-\left\langle \mathbf{\mathbf{Y}} ,H_{(M,\rho,s)}\vartriangleright
F_\mathbf{\mathbf{Y}}\right\rangle \notag \\
&=& \{F,H\}^{(CH)} + \{F,H\}^{(EM)}+\left\langle  F_\mathbf{A},H_M\vartriangleright \mathbf{A}\right\rangle \notag \\ &&-\left\langle H_\mathbf{A},F_{M}\vartriangleright \mathbf{A}\right\rangle
+
\left\langle \mathbf{\mathbf{Y}} ,F_{M} \vartriangleright H_ \mathbf{\mathbf{Y}}\right\rangle
-\left\langle \mathbf{\mathbf{Y}} ,H_{M}\vartriangleright F_\mathbf{\mathbf{Y}}\right\rangle,
\end{eqnarray}
with the variables $(M,\rho,s,\mathbf{A},\mathbf{Y})$ on the space $\mathfrak{k}^\ast\times T^\ast \mathfrak{U} $. Here, $\{F,H\}^{(CH)}$ is the hydrodynamics bracket in (\ref{eq.PB.CH}) and $\{F,H\}^{(EMc)}$ is the canonical electromagnetic bracket in (\ref{eq.PB.ED.Can}). In the calculation, we considered, for example, $F_{(M,\rho,s)}=(F_M,F_\rho,F_s)$ where $F_M$ is a vector field whereas $F_\rho$ and $F_s$ are real valued functions. This bracket is the same with the one given in \cite{holm1986hamiltonian}. To write the Poisson bracket in terms of the magnetic and electric fields $(\mathbf{B},\mathbf{E})$ we simply substitute (after identifying the one-form $\mathbf{A}$ with a vector field using the Euclidean metric on $\mathbb{R}^3$)
$\mathbf{B}=\nabla \times \mathbf{A}$ and $\mathbf{E}=-\mathbf{Y}$ and compute
\begin{eqnarray*}
\left\langle F_A, G_M\vartriangleright \mathbf{A} \right\rangle
=-\left\langle F_B,{G_M}  \vartriangleright  \mathbf{B} \right\rangle.
\end{eqnarray*}
We remark that, since $\mathbf{A}$ is a one-form, the action on the left hand side is the one in (\ref{Lie2}), since $\mathbf{B}$ is assumed to be a vector field,  the action on the right hand side is the one in (\ref{Lie1}). So, although the Poisson bracket
\begin{eqnarray} \label{eq.PB.EMHD}
\left\{F,H\right\}^{(EMHD)}= \{F,H\}^{(CH)} + \{F,H\}^{(EM)}-\left\langle  F_\mathbf{B},H_M\vartriangleright \mathbf{B} \right\rangle\nonumber\\+
 \left\langle H_\mathbf{B},F_{M}\vartriangleright \mathbf{B}\right\rangle
+\left\langle \mathbf{\mathbf{E}} ,F_{M} \vartriangleright H_ \mathbf{\mathbf{E}}\right\rangle
-\left\langle \mathbf{\mathbf{E}} ,H_{M}\vartriangleright F_\mathbf{\mathbf{E}}\right\rangle,
\end{eqnarray}
with the variables $(M,\rho,s,\mathbf{B},\mathbf{E})$, looks similarly as bracket (\ref{EMHDc}), in a local chart they will be different. Note also that, in (\ref{eq.PB.EMHD}) we substitute $\{F,H\}^{(EM)}$ the EM-bracket depending on the variables $(\mathbf{E,B})$, presented in (\ref{eq.PB.ED}), instead of the bracket $\{F,H\}^{(EMc)}$. Bracket \eqref{eq.PB.EMHD} is equivalent to bracket (14) of \cite{holm1986hamiltonian} when taking the displacement field $\mathbf{D}=\eps_0 \mathbf{E}$. Indeed, it can be then rewritten explicitly as
\begin{eqnarray} \label{eq.PB.EMHD--}
\left\{F,H\right\}^{(EMHD)}&=&\{F,H\}^{(CH)} + \{F,H\}^{(EM)} \notag
\\&+&\int d\rr \mathbf{E}\cdot[ (H_M\cdot\nabla) F_E -(F_M\cdot\nabla) H_E] \notag
\\ &+&\int d\rr \frac{z e}{m\eps_0} \rho(F_M \cdot H_E -H_M \cdot F_E) \notag
\\&+&\int d\rr F_M\cdot (\mathbf{E} \cdot \nabla) H_E - H_M\cdot (\mathbf{E} \cdot \nabla) F_E
\notag
\\&+&\int d\rr \mathbf{B}\cdot[ (H_M\cdot\nabla) F_B -(F_M\cdot\nabla) H_B]
\notag
\\&+&\int d\rr F_M\cdot (\mathbf{B} \cdot \nabla) H_B - H_M\cdot (\mathbf{B} \cdot \nabla) F_B,
\end{eqnarray}
where we have taken $H_\mathbf{E}$ as a one-form, we have employed the action in (\ref{Lie2}) while defining $ F_{M} \vartriangleright H_ \mathbf{E}$, and we have substituted two of the Maxwell's equations $\nabla \cdot \mathbf{E}=\frac{ze\rho}{m}$ and $\nabla \cdot \mathbf{B}=0$. In order to write this bracket in terms of velocities instead of the total momenta, we need to introduce a Legendre transformation between the momentum $M$ and the velocity $\mathbf{v}$. It should be borne in mind that the momenta $M$ in this bracket is in form $M=u +\eps_0\mathbf{E}\times \mathbf{B}$.

An alternative way to couple the EM-field with the hydrodynamics can be achieved by the projecting (KED) Poisson bracket presented in Eq.(\ref{eq.PB.KED}) via the following projections
\begin{subequations}
\begin{eqnarray}
\rho(\rr_a)&=&\int d\rr\int d\pp m f(\rr,\pp)\delta(\rr-\rr_a) \\
 u_i(\rr_a)&=&\int d\rr\int d\pp p_i f(\rr,\pp)\delta(\rr-\rr_a)\\
 s(\rr_a)&=&\int d\rr\int d\pp \sigma(f(\rr,\pp))\delta(\rr-\rr_a)\\
 \EE(\rr) &=& \EE(\rr)\\
 \BB(\rr) &=& \BB(\rr),
\end{eqnarray}
\end{subequations}
where the first three is called in the literature as the plasma-to-fluid map \cite{Marsden-Ratiu}. This way we obtain
\begin{eqnarray}\label{eq.PB.EMHD.alt}
 \{F,H\}^{(EMHD')}&=&\{F,H\}^{(CH)} + \int d \rr \frac{1}{\eps_0}\left(F_{\EE} \cdot (\nabla\times H_{\BB})-H_{\EE} \cdot (\nabla\times F_{\BB})\right)\nonumber\\
 &&+ \int d \rr \frac{z e}{m \eps_0} \rho \left(F_{u} \cdot H_\EE - H_{u} \cdot F_\EE\right)+\nonumber\\
 &&+ \int d \rr \frac{z e}{m} \rho \BB \cdot (F_{u} \times H_{u}),
\end{eqnarray}
governing the evolution of electromagnetohydrodynamics in variables $(\rho, u, s, \mathbf{E}, \mathbf{B})$. After transforming this bracket to variables $(\rho, M, s, \mathbf{E}, \mathbf{B})$, i.e. from hydrodynamic momentum to total momentum, the bracket becomes\footnote{Hand-written notes are available upon personal request to the corresponding author. The calculation were checked using the automated Poisson bracket manipulation program \cite{Kroeger2010}.} bracket \eqref{eq.PB.EMHD}. Both brackets are thus equivalent, only expressed in different variables.


\subsection{Binary electromagnetohydrodynamics}
Consider a binary fluid interacting with electromagnetic field. The Poisson bracket governing reversible evolution of these state variables can be obtained by projecting bracket \eqref{eq.PB.BKED} to state variables
\begin{subequations}
\begin{eqnarray}
\rho_1(\rr_a)&=&\int d\rr \int d\pp m f_1(\rr,\pp)\delta(\rr-\rr_a),\\
\rho_2(\rr_a)&=&\int d\rr \int d\pp m f_2(\rr,\pp)\delta(\rr-\rr_a),\\
u_i^1(\rr_a)&=&\int d\rr\int d\pp p_i f_1(\rr,\pp)\delta(\rr-\rr_a),\\
u_i^2(\rr_a)&=&\int d\rr\int d\pp p_i f_2(\rr,\pp)\delta(\rr-\rr_a),\\
 s_1(\rr_a)&=&\int d\rr\int d\pp \sigma_1(f_1(\rr,\pp))\delta(\rr-\rr_a),\\
 s_2(\rr_a)&=&\int d\rr\int d\pp \sigma_2(f_2(\rr,\pp))\delta(\rr-\rr_a),\\
s_2 &=& \int d \pp \sigma_2(f_2(\rr,\pp))
\end{eqnarray}
and $\EE$ and $\BB$. The projection results in
\end{subequations}
\begin{eqnarray}
 \{F,H\}^{(BEMHD)}&=&\{F,H\}^{(BH)} + \int d \rr \frac{1}{\eps_0}\left(F_{\EE} \cdot (\nabla\times H_{\BB})-H_{\EE} \cdot (\nabla\times F_{\BB})\right)+\nonumber\\
 &&+\sum_{\alpha=1}^2 \int d \rr \frac{z_\alpha e}{m_\alpha \eps_0} \rho_\alpha \left(F_{u^\alpha} \cdot H_\EE - H_{u^\alpha} \cdot F_\EE\right)+\nonumber\\
 &&+\sum_{\alpha=1}^2 \int d \rr \frac{z_\alpha e}{m_\alpha} \rho_\alpha \BB \cdot (F_{u^\alpha} \times H_{u^\alpha})
\end{eqnarray}
where $\eps_0$ is permittivity of vacuum, $z_\alpha$ is number of elementary charges per particle of species $\alpha$, $e$ is elementary charge and $m_\alpha$ is mass of a particle of species $\alpha$. This bracket has already been proposed in \cite{Spencer-Kaufman} and was shown to be implied by the canonical Poisson bracket of particle momenta and positions and electric intensity and vector potential in \cite{Holm1983}.

The evolution equations are then given by
\begin{subequations}
\begin{eqnarray}
\frac{\partial \rho_\alpha}{\partial t} &=& -\partial_i \left( \rho_\alpha H_{M^\alpha_i} \right)\\
\frac{\partial u^\alpha_i}{\partial t} &=& -\rho_\alpha \partial_i H_{\rho_\alpha} - u^\alpha_j \partial_i H_{u^\alpha_j} - s_\alpha H_{s_\alpha} - \partial_j (u^\alpha_i  H_{u^\alpha_j}) +\nonumber\\
&&+\frac{z_\alpha e}{m_\alpha}\rho_\alpha \left(\frac{1}{\eps_0}H_{E_i} +  \eps_{ijk}H_{u^\alpha_j}B_k\right)\\
\frac{\partial s_\alpha}{\partial t} &=& -\partial_i \left(s_\alpha H_{u^\alpha_i}\right)\\
\frac{\partial E_i}{\partial t} &=& \frac{1}{\eps_0} \eps_{ijk}\partial_j H_{B_k} - \sum_{\alpha = 1}^2 \frac{z_\alpha e}{m_\alpha \eps_0} \rho_\alpha H_{u^\alpha_i}\\
\frac{\partial B_i }{\partial t} &=& -\frac{1}{\eps_0}\eps_{ijk} \partial_j H_{E_k}
\end{eqnarray}
These evolution equations are also equipped with two constraints imposed on the Poisson bracket
\begin{eqnarray}
 \dive \BB &=& 0 \\
 \dive \EE &=& \frac{1}{\eps_0}\sum_{\alpha=1}^2 \frac{z_\alpha e}{m_\alpha}\rho_\alpha,
\end{eqnarray}
which follow from gauge invariance of the electromagnetic fields as shown in \cite{Marsden1982}.
\end{subequations}

Energy could be chosen for example as
\begin{equation}
 H = \int d \rr \sum_{\alpha = 1}^2\left(\frac{(u^\alpha)^2}{2\rho_\alpha} + \eps_\alpha(\rho_\alpha, s_\alpha)\right) + \eps_I(\rho_1, \rho_2, s_1, s_2, \EE, \BB) + \frac{1}{2}\eps_0 \EE^2 + \frac{1}{2\mu_0}\BB^2
\end{equation}
where $\mu_0$ is vacuum permeability and $\eps_I$ is an interaction energy among the species and the electromagnetic field. Note that the terms $\eps_\alpha$ form partial pressures in the partial momentum evolution equations while the interaction term gives rise to reversible momentum exchange as commented in \cite{Pavelka-IJES}.

\subsection{Classical binary electromagnetohydrodynamics}
Poisson bracket \eqref{eq.PB.BKED} can be further reduced to one-temperature-one-momentum fluid, given by state variables $(M,\rho_1, \rho_2,s,\EE, \BB)$, by projection
\begin{equation}
 u = u^1 + u^2, s = s_1 + s_2.
\end{equation}
The Poisson bracket is then
\begin{eqnarray}
 \{F,G\}^{(CBEMHD)} &=& \{F,G\}^{(CBH)} + \{F,G\}^{(EM)}+\nonumber\\
 &&+\sum_{\alpha = 1}^2 \intd\rr \frac{z_\alpha e}{m_\alpha \eps_0} \rho_\alpha \left(F_u \cdot G_\EE - G_u \cdot F_\EE\right)+\nonumber\\
 &&+\sum_{\alpha = 1}^2 \intd\rr \frac{z_\alpha e}{m_\alpha} \rho_\alpha \BB \cdot \left(F_u \times G_u\right).
\end{eqnarray}

\subsection{Magnetohydrodynamics}
In magnetohydrodynamics electric intensity no longer plays the role of state variable \cite{goedbloed2004principles}. The direct way to arrive at the magnetohydrodynamics bracket, we may consider the (EMHD) bracket in \eqref{eq.PB.EMHD--} and consider the particular case of that $F_E$ and $H_E$ are zero, so that we have
\begin{eqnarray} \label{eq.PB.MHD}
\left\{F,H\right\}^{(MHD)}&=&\{F,H\}^{(CH)} \notag
\\&+&\int d\rr \mathbf{B}\cdot[ (H_M\cdot\nabla) F_B -(F_M\cdot\nabla) H_B]
\notag
\\&+&\int d\rr F_M\cdot (\mathbf{B} \cdot \nabla) H_B - H_M\cdot (\mathbf{B} \cdot \nabla) F_B.
\end{eqnarray}
This is Lie Poisson bracket on the dual space presented implicitly in \eqref{LiePoissonongh}.
We refer \cite{Holm1983,Holm-Kupershmidt1983-MHD,MaWeRaScSp83,Morrison-Greene1980,Morrison-Greene-erratum} for more details on this bracket.

Evolution equations are then given by a choice of energy $E$ and by rewriting the bracket as
\begin{equation}
 \{F,E\}^{(MHD)} = \intd\rr
 \frac{\partial F}{\partial \rho} \frac{\partial \rho}{\partial t}
 +\frac{\partial F}{\partial M} \cdot \frac{\partial M}{\partial t}
 +\frac{\partial F}{\partial s} \frac{\partial s}{\partial t}
 +\frac{\partial F}{\partial \BB} \cdot \frac{\partial \BB}{\partial t}.
\end{equation}
Specifying energy as
\begin{equation}
 H = \intd\rr \frac{M^2}{2\rho} + \eps(\rho,s) + \frac{1}{2\mu_0} \BB^2
\end{equation}
then leads to evolution equations
\begin{subequations}\label{eq.evo.MHD}
\begin{eqnarray}
 \frac{\partial \rho}{\partial t} &=&-\nabla\cdot M,\\
 \frac{\partial M}{\partial t} &=& -\nabla p - \nabla\cdot\left(\frac{M\otimes M}{\rho}\right) + \frac{1}{\mu_0}(\nabla \times \BB)\times\BB,\\
 \frac{\partial s}{\partial t} &=& \nabla\cdot\left(s\frac{M}{\rho}\right),\\
 \frac{\partial \BB}{\partial t} &=& -\nabla\times \left(M\times \frac{M}{\rho}\right)
\end{eqnarray}
where pressure $p$ is given by
\begin{equation}
 p = -\eps + \rho \eps_\rho + s \eps_s.
\end{equation}
\end{subequations}
Equations \eqref{eq.evo.MHD} represent the reversible part of the standard magnetohydrodynamic equations.

\subsection{Electrohydrodynamics}
Now, we assume that the function(al)s on the electromagnetohydrodynamics bracket (\ref{eq.PB.EMHD}) does not depend on the magnetic field $\mathbf{B}$. Such an assumption is also meaningful in the geometric setting of the Poisson structure. For this case, it is not possible to direct application of Lie-Poisson reduction, instead we refer the Poisson structure (\ref{eq.PBh*V}). So that, we have that
\begin{eqnarray*} \label{eq.PB.EHD}
\left\{A,B\right\}^{(EHD)}= \{A,B\}^{(CH)}+\left\langle \mathbf{\mathbf{E}} ,A_{M} \vartriangleright B_ \mathbf{\mathbf{E}}\right\rangle-\left\langle \mathbf{\mathbf{E}} ,B_{M}\vartriangleright A_\mathbf{\mathbf{E}}\right\rangle,
\end{eqnarray*}
with variables $(M,\rho,s,\mathbf{E})$. Explicitly, we have that
\begin{eqnarray} \label{eq.PB.EHD-}
\left\{F,H\right\}^{(EMHD)}&=&\{F,H\}^{(CH)} \notag
\\&+&\int d\rr \mathbf{E}\cdot[ (H_M\cdot\nabla) F_E -(F_M\cdot\nabla) H_E] + \rho(F_M \cdot H_E -H_M \cdot F_E) \notag
\\&+&\int d\rr F_M\cdot (\mathbf{E} \cdot \nabla) H_E - H_M\cdot (\mathbf{E} \cdot \nabla) F_E.
\end{eqnarray}
See \cite{holm1986hamiltonian,Holm-Kupershmidt1983-MHD} for more details on the electrohydrodynamics Poisson bracket.

\section{Onsager-Casimir reciprocal relations}\label{sec.OCRR}
Onsager-Casimir reciprocal relations \cite{Casimir1945,Onsager1930,Onsager1931}, comprehensively reviewed in \cite{dGM}, say that variables with the same parities with respect to time-reversal transformation are coupled by a symmetric matrix while variables with opposite parities are coupled by an antisymmetric matrix. Within the GENERIC framework \cite{GO,OG} the antisymmetric coupling is given by the Hamiltonian part of the evolution equations, i.e. by the Poisson bracket, see \cite{PRE2014}. Let us now have a look at how such antisymmetric coupling provided by a Poisson bracket works in the presence of magnetic field.

The time-reversal transformation is an operation which inverts velocities of all particles and magnetic field. Physical quantities can be then even with respect to time-reversal (do not change sign, e.g. density, energy density), odd (do change sign, e.g. momentum) or without any definite parity (e.g. the Boltzmann distribution function). Even quantities have parity equal to $1$ while odd quantities have parity equal to $-1$, i.e. when denoting time-reversal by $\mathbf{I}$,
\begin{eqnarray}
\mathbf{I}(x) = x \rightarrow \mathcal{P}(x) = 1 \mbox{ and }
\mathbf{I}(x) = -x \rightarrow \mathcal{P}(x) = -1,
\end{eqnarray}
see \cite{PRE2014}.

Consider now Poisson bracket \eqref{eq.PB.EMHD.alt}. The first line of the bracket is indeed compatible with the Onsager-Casimir reciprocal relations as was shown in \cite{HCO,PRE2014}. The second line provides coupling between momentum, which is an odd variable, and electric intensity, which is even. That line thus provides antisymmetric coupling between variables with different parities, which is compatible with the meaning of the Casimir coupling.

The third line provides coupling of momentum with itself. At first sight it might seem that such coupling should not be antisymmetric because of the same parity of the coupled variables. However, Onsager-Casimir reciprocal relations become more subtle in the presence of magnetic field. Symmetry or antisymmetry of the coupling matrices are still given by parities of the coupled variables, but $\BB$ is replaced by $-\BB$ in the symmetric or antisymmetric counterpart. In particular, variables with the same parities exhibit symmetric coupling, but $\BB$ is replaced by $-\BB$ in the symmetric counterpart.

The third line of bracket \eqref{eq.PB.EMHD} can be rewritten as
\begin{eqnarray}
 \int d\rr_1\int d\rr_2 \frac{z e}{m}\rho(\rr_1)\delta(\rr_1-\rr_2)\left(B_x(\rr_1)F_{u_y}(\rr_1)G_{u_z}(\rr_2)-B_y(\rr_1)F_{u_z}(\rr_1)G_{u_y}(\rr_2)\right)\nonumber\\
 +\int d\rr_1\int d\rr_2 \frac{z e}{m}\rho(\rr_1)\delta(\rr_1-\rr_2)\left(B_y(\rr_1)F_{u_z}(\rr_1)G_{u_x}(\rr_2)-B_y(\rr_1)F_{u_x}(\rr_1)G_{u_z}(\rr_2)\right)\nonumber\\
 +\int d\rr_1\int d\rr_2 \frac{z e}{m}\rho(\rr_1)\delta(\rr_1-\rr_2)\left(B_z(\rr_1)F_{u_x}(\rr_1)G_{u_y}(\rr_2)-B_z(\rr_1)F_{u_y}(\rr_1)G_{u_x}(\rr_2)\right),
\end{eqnarray}
which can be interpreted as that when replacing $\BB$ by $-\BB$ in the coefficient in front of for example $F_{u_y} G_{u_z}$, we obtain the coefficient in front of $F_{u_z} G_{u_y}$. Such coupling can be thus regarded as effectively symmetric, which corresponds to the same parity of the coupled variables, just with $\BB$ replaced by $-\BB$. In other words, coupling between momentum and momentum provided by bracket \eqref{eq.PB.EMHD} is symmetric with respect to simultaneous transposition and time-reversal transformation.

In summary, Onsager-Casimir reciprocal relations in the presence of magnetic field are compatible with the coupling provided by Poisson bracket \eqref{eq.PB.EMHD} provided transposition is carried out simultaneously with inversion of odd quantities. Moreover, the same holds true about bracket \eqref{eq.PB.MHD} repeating the above argumentation. This is also the content of the bare and dressed symmetries introduced in \cite{HCO}.

Note also that reversibility of the Hamiltonian evolution with respect to time-reversal requires that parity of coefficients in front of variables with the same parity is odd while parity of coefficients in front of variables with opposite parities is even, see \cite{PRE2014}. This is obviously fulfilled in brackets \eqref{eq.PB.EMHD} and \eqref{eq.PB.MHD}.

Let us now summarize the general formulation of Onsager-Casimir reciprocal relations by combining results from \cite{HCO} and \cite{PRE2014}. Each evolution equation can be split into its reversible and irreversible part. Assuming that the evolution equation is in the GENERIC form \cite{GO,OG}, the reversible part is given by a Poisson bracket while the irreversible part is given by a dissipation potential. When sufficiently near to thermodynamic equilibrium (when the thermodynamic forces are small enough), the convex dissipation potential can be approximated by a quadratic function, and the irreversible evolution can be expressed in terms of a dissipative bracket \cite{HCO}. The evolution equation than has the form
\begin{equation}
 \dot{x}^i = L^{ij}\Phi_{x_j} + M^{ij} \Phi_{x^j}
\end{equation}
where $\xx$ is the vector of state variables, $\mathbf{L}$ is the Poisson bivector giving a corresponding Poisson bracket $\langle d A, \mathbf{L}\cdot dB\rangle$, $\Phi$ is free energy and $\mathbf{M}$ is the dissipative bracket. Poisson bivector matrix $\mathbf{L}$ is antisymmetric due to antisymmetry of the corresponding Poisson bracket while the dissipative matrix is symmetric because it is equal to second differential of the dissipation potential (by equality of second mixed derivatives). Antisymmetry of $\mathbf{L}$ and symmetry of $\mathbf{M}$ is the bare symmetry introduced in \cite{HCO}.

It was shown in \cite{PRE2014} that the entries of the Poisson bivector matrix have opposite parities than the variables they are coupling while entries of the dissipative bracket have the same parities, namely
\begin{equation}
 \mathcal{P}(L^{ij}) = - \mathcal{P}(x^i) \mathcal{P}(x^j)
 \mbox{ and }\mathcal{P}(M^{ij}) = \mathcal{P}(x^i) \mathcal{P}(x^j)
\end{equation}
where $\mathcal{P}(x)= 1$ for $x$ even with respect to time-reversal (e.g. density, energy density, entropy density, electric intensity) while $\mathcal{P}(x)=-1$ for odd variables (e.g. momentum density, magnetic induction).

Consider now two variables with the same parity. The Poisson bivector provides no or antisymmetric coupling between them. However, parity of the entry of the Poisson bivector matrix providing that coupling is $-1$, and thus when the time-reversal transformation is applied (odd variables change signs), the coupling becomes effectively symmetric. That is the dressed symmetry proposed in \cite{HCO}. Similarly, the dissipative matrix provides symmetric coupling between the variables, and its entries are not altered by time-reversal so that the coupling remains symmetric. Variables with the same parity are then effectively coupled by a symmetric matrix. The matrix is given by the symmetric part of the Poisson bivector with respect to simultaneous transposition and time-reversal and by the dissipative matrix (which is symmetric and independent of time-reversal by construction).

Consider now two variables with opposite parities. The Poisson bivector again provides antisymmetric coupling between them. Since the parity of the particular entry of the matrix is $+1$, the entry does not change its sign when time-reversal is applied. Therefore, the coupling remains antisymmetric even when time-reversal is applied. Similarly, the dissipative matrix provides symmetric coupling between the variables, but the particular entry of the matrix is odd with respect to time-reversal. Therefore, upon time-reversal the coupling becomes effectively antisymmetric. This is again the dressed symmetry introduced in \cite{HCO}.

In summary, let us denote the sum of the two matrices as $K^{ij}(\xx) = L^{ij}(\xx)+M^{ij}(\xx)$. Matrix $\mathbf{K}$ is the matrix of phenomenological coefficients. The Onsager-Casimir reciprocal relations are then expressed by
\begin{equation}\label{eq.OCRR}
 K^{ij}(\xx) = \mathcal{P}(x^i)\mathcal{P}(x^j) K^{ji}(\mathbf{I}(\xx)).
\end{equation}
Note that not only magnetic field is inverted during the time-reversal (as in the original Onsager-Casimir reciprocal relations, see e.g. \cite{dGM}), but also all other odd variables are inverted. Equation \eqref{eq.OCRR} thus represents a generalization of Onsager-Casimir reciprocal relations.

\section{Conclusion}
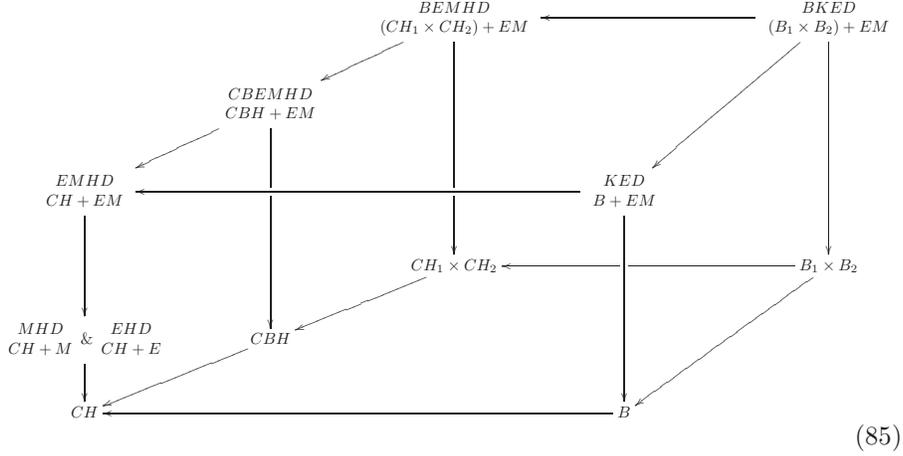
\begin{figure}[ht]
\begin{equation}
\resizebox{\textwidth}{!}{
\xymatrix{
&&{\begin{array}{c}BEMHD \\ (CH_1\times CH_2)+ EM \end{array}} \ar[ld]\ar[ddd]|!{[d];[dd]}\hole&&&
{\begin{array}{c}BKED \\(B_1 \times B_2)+ EM \end{array}}
\ar[ddd]\ar[lll]\ar[lldd]
\\
&
{\begin{array}{c}CBEMHD \\CBH+ EM \end{array}}
\ar[ld]\ar[ddd]|!{[d];[d]}\hole&&&
\\
{\begin{array}{c}EMHD \\CH+ EM \end{array}}
\ar[dd]&&&
{\begin{array}{c}KED \\B+ EM \end{array}}
\ar[ddd]\ar[lll]&
\\
& 
&CH_1\times CH_2\ar[ld]&&&B_1 \times B_2\ar[lldd]\ar[lll]|!{[l];[ll]}\hole
\\
{\begin{array}{c}MHD \\CH+M \end{array} \& \begin{array}{c}EHD \\CH+E \end{array}}
\ar[d]&CBH\ar[ld]&&&&
\\
CH&&&B\ar[lll]&
}}
\end{equation}
\caption{\label{fig.levels}Considered levels of description. The acronyms stand for (form right top to left bottom): binary kinetic electrodynamics, binary electromagnetohydrodynamics, classical binary electromagnetohydrodynamics, kinetic electrodynamics, electromagnetohydrodynamics, binary kinetic theory, binary hydrodynamics, classical binary hydrodynamics, magnetohydrodynamics and electrohydrodynamics, kinetic theory and classical hydrodynamics. The ground floor contains classical hydrodynamics and Boltzmann equation (kinetic theory), the first floor contains systems obtained by the semidirect products of the one in the ground floor and electromagnetic theory. The entresol contains projections from the first floor down.
}
\end{figure}

Evolution equations of mesoscopic models in physics and engineering can be split into reversible and irreversible part. The reversible part is then often generated by a Poisson bracket and an energy (Hamiltonian function). Such a point of view on evolution equations has become important in non-equilibrium thermodynamics, but to promote it in different fields of physics and engineering it is necessary to present construction of Poisson brackets used in practical situations in an uncluttered and general way. The main goal of this paper is to elucidate how to construct Poisson brackets coupling matter (in particular mixtures of fluids) with electrodynamics.

Poisson brackets are objects extensively studied in differential geometry, and they are often implied by a Lie group describing evolution of a dynamical system. When more details of the system are taken into account, for example its electromagnetic behavior, the Lie group has to be coupled with the Lie group expressing evolution of the electromagnetic field. We try to explain in a gentle way how to construct the Poisson brackets from the respective Lie groups and how to couple them by means of direct product (no interaction between the groups), semidirect product (one group is Lie-dragged by the other) and matched pairs (both of groups are dragged by each other).

On an abstract level, we have presented in Sec.(\ref{CTHS}) a purely geometrical way to match (couple) two canonical Poisson brackets as well as to match two Lie-Poisson brackets. We have started to this section with recalling basic notions related with the matched pair of Lie groups. A matched pair Lie group is a union of two of its Lie subgroups which are mutual interaction. The cotangent bundle of a matched pair Lie group carries a canonical Poisson structure (\ref{PoissononT*GH2}), which we have proposed as the correct coupling of the canonical Poisson structures on the cotangent bundles of the constitutive Lie groups. Accordingly, the associated matched pair Lie algebra is the direct sum of two Lie subalgebras in mutual interaction. The dual space carries the matched pair Lie-Poisson bracket (\ref{LiePoissonongh}) which we have proposed as the correct way to couple two Lie-Poisson systems in mutual interaction. After presenting these theoretical frameworks, we have devoted the rest of this study for presenting some of the previously known results in terms of the general framework of the matched dynamics.

In particular, coupling between electromagnetic field and transport of matter is considered on different levels of description (differing by the amount of detail expressed by the chosen state variables). The levels of description are summarized in Fig. \ref{fig.levels}. A wide range of levels of description is covered, going from kinetic theory to hydrodynamics including mixtures on both the kinetic and hydrodynamic level. Let us now briefly summarize how the particular levels are constructed. The Poisson bracket of electrodynamics can be constructed from the canonical Poisson bracket on a cotangent bundle as in Sec. \ref{sec.ED}, and the Poisson bracket of kinetic theory is given as the implied bracket on Lie-algebra dual of the group of canonical transformations, and it is thus determined by the canonical Poisson bracket of classical mechanics. Coupling between kinetic theory and electrodynamics is demonstrated in Sec. \ref{sec.KED}.

Kinetic theory can be projected to classical hydrodynamics, see Sec. \ref{sec.HD}, where the construction of classical hydrodynamics by means of a semidirect product (density and entropy density dragged by the momentum field) is demonstrated as well. Similarly, kinetic electrodynamics can be projected to electromagnetohydrodynamics. The latter level of description can be also conveniently constructed by means of semidirect product coupling as shown in Sec. \ref{sec.EMHD}. Electromagnetohydrodynamics can be then projected to electrohydrodynamics (dropping magnetic induction) or magnetohydrodynamics (dropping electric intensity).

Kinetic theory can be easily extended to mixtures (considered only binary mixtures for simplicity of notation) by means of a direct product, and such a binary kinetic theory can be then coupled with electrodynamics as in the case of the one-species kinetic theory, see Sec. \ref{sec.BKED}. Similarly, classical hydrodynamics can be extended to binary mixtures in the context of Extended Irreversible Thermodynamics \cite{Jou-EIT} by means of direct product of the respective Lie group, but the binary hydrodynamics can be also obtained by projection from binary kinetic theory. Binary electromagnetohydrodynamics (in the context of Classical and Extended Irreversible Thermodynamics) can be then obtained by projection from binary kinetic electrodynamics.

In Sec. \ref{sec.OCRR} we discuss the meaning of Onsager-Casimir reciprocal relations (OCRR) in the context of Hamiltonian evolution containing electromagnetic field. The Poisson brackets lead only to antisymmetric coupling, but by inverting the field of magnetic induction (as in the standard meaning of OCRR) we obtain an effectively symmetric coupling as required by OCRR when coupling state variables with the same parities with respect to time reversal (e.g. magnetic induction with itself). Onsager-Casimir reciprocal relations are thus compatible with the Hamiltonian coupling between electromagnetic field and matter.

In summary, we discuss constructions of Poisson brackets expressing evolution of electromagnetic field and matter. Some of the brackets are constructed by means of geometrical coupling (direct, semidirect or matched-pair products) and some are given by projections from more detailed Poisson brackets. In particular, mixtures on the kinetic and hydrodynamic levels are coupled with electrodynamics. The meaning of Onsager-Casimir reciprocal relations in the context of electrodynamics is also discussed.

\section*{Acknowledgement}
M. P. is grateful to professor František Maršík for his generous support and for revealing the world of thermodynamics. O.E. is grateful to professor Hasan G\"{u}mral for the enlightening discussions on the Lie-Poisson dynamics and to professor Serkan S\"{u}tl\"{u} for the invaluable comments on the applications of the matched pair dynamics in case of the field theories.

This project was supported by Natural Sciences and Engineering Research Council of Canada (NSERC).

The work was partially developed within the POLYMEM project, reg. no CZ.1.07/2.3.00/20.0107, that is co-funded from the European Social Fund (ESF) in the Czech Republic: ``Education for Competitiveness Operational Programme'', from the CENTEM project, reg. no. CZ.1.05/2.1.00/03.0088, co-funded by the ERDF as part of the Ministry of Education, Youth and Sports OP RDI programme and, in the follow-up sustainability stage, supported through CENTEM PLUS (LO1402) by financial means from the Ministry of Education, Youth and Sports under the ``National Sustainability Programme I''. 

Further, this work was also supported by Charles University in Prague, project GA UK No 70515, and by Czech Science Foundation, project no.  17-15498Y.

\bibliographystyle{plain}
\bibliography{library}

\begin{thebibliography}{10}

\bibitem{Abraham-Marsden}
Ralph Abraham and Jerrold~E. Marsden.
\newblock {\em Foundations of Mechanics}.
\newblock AMS Chelsea publishing. AMS Chelsea Pub./American Mathematical
  Society, 1978.

\bibitem{abraham2012manifolds}
Ralph Abraham, Jerrold~E Marsden, and Tudor Ratiu.
\newblock {\em Manifolds, tensor analysis, and applications}, volume~75.
\newblock Springer Science \& Business Media, 2012.

\bibitem{agrachev2013control}
Andrei~A Agrachev and Yuri Sachkov.
\newblock {\em Control theory from the geometric viewpoint}, volume~87.
\newblock Springer Science \& Business Media, 2013.

\bibitem{Arnoldbook}
V.~I. Arnold.
\newblock {\em Mathematical methods of classical mechanics}.
\newblock Springer, New York, 1989.

\bibitem{Arnold}
V.I. Arnold.
\newblock Sur la g\'{e}ometrie diff\'{e}rentielle des groupes de lie de
  dimension infini et ses applications dans l'hydrodynamique des fluides
  parfaits.
\newblock {\em Annales de l'institut Fourier}, 16(1):319--361, 1966.

\bibitem{arnold1999topological}
Vladimir~I Arnold and Boris~A Khesin.
\newblock {\em Topological methods in hydrodynamics}, volume 125.
\newblock Springer Science \& Business Media, 1999.

\bibitem{bloch1996nonholonomic}
Anthony~M Bloch, PS~Krishnaprasad, Jerrold~E Marsden, and Richard~M Murray.
\newblock Nonholonomic mechanical systems with symmetry.
\newblock {\em Archive for Rational Mechanics and Analysis}, 136(1):21--99,
  1996.

\bibitem{bruveris2011momentum}
Martins Bruveris, Fran{\c{c}}ois Gay-Balmaz, Darryl~D Holm, and Tudor~S Ratiu.
\newblock The momentum map representation of images.
\newblock {\em Journal of nonlinear science}, 21(1):115--150, 2011.

\bibitem{bruveris2015geometry}
Martins Bruveris and Darryl~D Holm.
\newblock Geometry of image registration: The diffeomorphism group and momentum
  maps.
\newblock In {\em Geometry, Mechanics, and Dynamics}, pages 19--56. Springer,
  2015.

\bibitem{FB-ADL:04}
Francesco Bullo and Andrew~D. Lewis.
\newblock {\em Geometric Control of Mechanical Systems}, volume~49 of {\em
  Texts in Applied Mathematics}.
\newblock Springer Verlag, New York-Heidelberg-Berlin, 2004.

\bibitem{cartan2012differential}
Henri Cartan.
\newblock {\em Differential forms}.
\newblock Courier Corporation, 2012.

\bibitem{Casimir1945}
H.~B.~G. Casimir.
\newblock On {O}nsager's principle of microscopic reversibility.
\newblock {\em Rev. Mod. Phys.}, 17:343--350, Apr 1945.

\bibitem{cendra1998maxwell}
Hern{\'a}n Cendra, Darryl~D Holm, Mark~JW Hoyle, and Jerrold~E Marsden.
\newblock The maxwell--vlasov equations in euler--poincar{\'e} form.
\newblock {\em Journal of Mathematical Physics}, 39(6):3138--3157, 1998.

\bibitem{dGM}
S.~R. de~Groot and P.~Mazur.
\newblock {\em Non-equilibrium Thermodynamics}.
\newblock Dover Publications, New York, 1984.

\bibitem{Di64}
P.A.M. Dirac.
\newblock {\em Lectures in quantum mechanics}.
\newblock Yeshiva University, 1964.

\bibitem{ebinmarsden}
D.G. Ebin and J.E. Marsden.
\newblock Groups of diffeomorphisms and the motion of an incompressible fluid.
\newblock {\em Annals of Mathematics}, 92(1):102--163, 1970.

\bibitem{ellis2010symmetry}
David~CP Ellis, Fran{\c{c}}ois Gay-Balmaz, Darryl~D Holm, Vakhtang Putkaradze,
  and Tudor~S Ratiu.
\newblock Symmetry reduced dynamics of charged molecular strands.
\newblock {\em Archive for rational mechanics and analysis}, 197(3):811--902,
  2010.

\bibitem{esen2012geometry}
O.~Esen and H.~G{\"u}mral.
\newblock Geometry of plasma dynamics ii: Lie algebra of hamiltonian vector
  fields.
\newblock {\em Journal of Geometric Mechanics}, 4(3), 2012.

\bibitem{esen2015lagrangian}
O.~Esen and S.~S{\"u}tl{\"u}.
\newblock Lagrangian dynamics on matched pairs.
\newblock {\em arXiv preprint arXiv:1512.06770}, 2015.

\bibitem{esen2016hamiltonian}
O.~Esen and S.~S{\"u}tl{\"u}.
\newblock Hamiltonian dynamics on matched pairs.
\newblock {\em arXiv preprint arXiv:1604.05130}, 2016.

\bibitem{esen2016vlasov}
{Esen, O., and S{\"u}tl{\"u}, S.}
\newblock Matched pairs decomposition of vlasov equation.
\newblock {\em In preperation}, 2016.

\bibitem{Fecko}
M.~Fecko.
\newblock {\em Differential Geometry and Lie Groups for Physicists}.
\newblock Cambridge University Press, 2006.

\bibitem{GaVi12}
F.~Gay-Balmaz and C.~Vizman.
\newblock Dual pairs in fluid dynamics.
\newblock {\em Annals of Global Analysis and Geometry}, 41(1):1--24, 2012.

\bibitem{goedbloed2004principles}
Johan~Peter Goedbloed and Stefaan Poedts.
\newblock {\em Principles of magnetohydrodynamics: with applications to
  laboratory and astrophysical plasmas}.
\newblock Cambridge university press, 2004.

\bibitem{Miroslav-turbulence}
M~Grmela, D~Jou, J~Casas-Vazquez, M~Bousmina, and G~Lebon.
\newblock Ensemble averaging in turbulence modelling.
\newblock {\em Physics Letters A}, 330(1-2):54--64, SEP 13 2004.

\bibitem{GO}
Miroslav Grmela and Hans~Christian \"{O}ttinger.
\newblock Dynamics and thermodynamics of complex fluids. {I}. {D}evelopment of
  a general formalism.
\newblock {\em Phys. Rev. E}, 56:6620--6632, Dec 1997.

\bibitem{gumral2010geometry}
Hasan G{\"u}mral.
\newblock Geometry of plasma dynamics. i. group of canonical diffeomorphisms.
\newblock {\em Journal of Mathematical Physics}, 51(8):083501, 2010.

\bibitem{holm1986hamiltonian}
Darryl~D Holm.
\newblock Hamiltonian dynamics of a charged fluid, including electro-and
  magnetohydrodynamics.
\newblock {\em Physics Letters A}, 114(3):137--141, 1986.

\bibitem{Holm1983}
Darryl~D. Holm and Boris~A. Kupershmidt.
\newblock Poisson brackets and clebsch representations for
  magnetohydrodynamics, multifluid plasmas, and elasticity.
\newblock {\em Physica D: Nonlinear Phenomena}, 6(3):347 -- 363, 1983.

\bibitem{Holm-Kupershmidt1983-MHD}
DD~Holm and BA~Kupershmidt.
\newblock {Noncanonical {H}amiltonian-formulation of ideal
  magnetohydrodynamics}.
\newblock {\em {Physica D}}, {7}({1-3}):{330--333}, {1983}.

\bibitem{hutter-plastic}
Markus H\"{u}tter and Bob Svendsen.
\newblock Thermodynamic model formulation for viscoplastic solids as general
  equations for non-equilibrium reversible-irreversible coupling.
\newblock {\em Continuum Mech. Thermodyn.}, 24:211--227, 2012.

\bibitem{Jou-EIT}
D.~Jou, J.~Casas-Vázquez, and G.~Lebon.
\newblock {\em Extended Irreversible Thermodynamics}.
\newblock Springer-Verlag, New York, 4th edition, 2010.

\bibitem{jurdjevic1997geometric}
Velimir Jurdjevic.
\newblock {\em Geometric control theory}.
\newblock Cambridge university press, 1997.

\bibitem{KoMiSl93}
I.~Kol\'{a}r, P.~W. Michor, and J.~Slov'{a}k.
\newblock {\em Natural operations in differential geometry}.
\newblock Springer-Verlag, Berlin, 1993.

\bibitem{Kroeger2010}
M.~Kroeger and M.~Huetter.
\newblock Automated symbolic calculations in nonequilibrium thermodynamics.
\newblock {\em Comput. Phys. Commun.}, 181:2149–2157, 2010.

\bibitem{Landau2}
L.D. Landau and E.~M. Lifshitz.
\newblock {\em The Classical Theory of Fields}.
\newblock Number v. 2 in Course of theoretical physics. Butterworth Heinemann,
  1975.

\bibitem{libermann2012symplectic}
Paulette Libermann and Charles-Michel Marle.
\newblock {\em Symplectic geometry and analytical mechanics}, volume~35.
\newblock Springer Science \& Business Media, 2012.

\bibitem{LuWein90}
J.-H. Lu and A.~Weinstein.
\newblock Poisson {L}ie groups, dressing transformations, and {B}ruhat
  decompositions.
\newblock {\em J. Differential Geom.}, 31(2):501--526, 1990.

\bibitem{Maji90-II}
S.~Majid.
\newblock Matched pairs of {L}ie groups associated to solutions of the
  {Y}ang-{B}axter equations.
\newblock {\em Pacific J. Math.}, 141(2):311--332, 1990.

\bibitem{Maji90}
S.~Majid.
\newblock Physics for algebraists: noncommutative and noncocommutative {H}opf
  algebras by a bicrossproduct construction.
\newblock {\em J. Algebra}, 130(1):17--64, 1990.

\bibitem{Majid-book}
S.~Majid.
\newblock {\em Foundations of quantum group theory}.
\newblock Cambridge University Press, Cambridge, 1995.

\bibitem{Marsden1982}
J.~E. Marsden and A.~Weinstein.
\newblock The {H}amiltonian structure of the {M}axwell-{V}lasov equations.
\newblock {\em Physica D}, pages 394--406, 1982.

\bibitem{Marsden-Ratiu}
J.E. Marsden and T.~S. Ratiu.
\newblock {\em Introduction to Mechanics and Symmetry}, volume Second edition
  of {\em Texts in Applied Mathematics 17}.
\newblock Springer-Verlag, New York, 1999.

\bibitem{MaWeRaScSp83}
J.E. Marsden, A.~Weinstein, T.~Ratiu, R.~Schmid, and R.G. Spencer.
\newblock Hamiltonian systems with symmetry, coadjoint orbits and plasma
  physics.
\newblock {\em Proceedings of the IUTAM-ISIMM symposium on modern developments
  in analytical mechanics 117, No. CAG-CONF-1983-001}, pages 289--340, 1983.

\bibitem{marsden1974reduction}
Jerrold Marsden and Alan Weinstein.
\newblock Reduction of symplectic manifolds with symmetry.
\newblock {\em Reports on mathematical physics}, 5(1):121--130, 1974.

\bibitem{marsden2007hamiltonian}
Jerrold~E Marsden, Gerard Misiolek, Juan-Pablo Ortega, Matthew Perlmutter, and
  Tudor~S Ratiu.
\newblock {\em Hamiltonian reduction by stages}.
\newblock 2007.

\bibitem{marsden1998symplectic}
Jerrold~E Marsden, Gerard Misio{\l}ek, Matthew Perlmutter, and Tudor~S Ratiu.
\newblock Symplectic reduction for semidirect products and central extensions.
\newblock {\em Differential Geometry and its Applications}, 9(1):173--212,
  1998.

\bibitem{marsden1984semidirect}
Jerrold~E Marsden, Tudor Ra{\c{t}}iu, and Alan Weinstein.
\newblock Semidirect products and reduction in mechanics.
\newblock {\em Transactions of the American Mathematical Society},
  281(1):147--177, 1984.

\bibitem{marsden1984reduction}
Jerrold~E Marsden, Tudor~S Ratiu, and Alan Weinstein.
\newblock Reduction and hamiltonian structures on duals of semidirect product
  lie algebras.
\newblock {\em Cont. Math. AMS}, 28:55--100, 1984.

\bibitem{meyer1973symmetries}
Kenneth~R Meyer.
\newblock Symmetries and integrals in mechanics.
\newblock {\em Dynamical systems}, pages 259--273, 1973.

\bibitem{Morrison-Greene1980}
Philip~J. Morrison and John~M. Greene.
\newblock Noncanonical hamiltonian density formulation of hydrodynamics and
  ideal magnetohydrodynamics.
\newblock {\em Phys. Rev. Lett.}, 45:790--794, Sep 1980.

\bibitem{Morrison-Greene-erratum}
Philip~J. Morrison and John~M. Greene.
\newblock Noncanonical hamiltonian density formulation of hydrodynamics and
  ideal magnetohydrodynamics.
\newblock {\em Phys. Rev. Lett.}, 48:569--569, Feb 1982.

\bibitem{nakahara2003geometry}
Mikio Nakahara.
\newblock {\em Geometry, topology and physics}.
\newblock CRC Press, 2003.

\bibitem{olver2000applications}
Peter~J Olver.
\newblock {\em Applications of Lie groups to differential equations}, volume
  107.
\newblock Springer Science \& Business Media, 2000.

\bibitem{Onsager1930}
Lars Onsager.
\newblock Reciprocal relations in irreversible processes. {I.}
\newblock {\em Phys. Rev.}, 37:405--426, Feb 1931.

\bibitem{Onsager1931}
Lars Onsager.
\newblock Reciprocal relations in irreversible processes. ii.
\newblock {\em Phys. Rev.}, 38:2265--2279, Dec 1931.

\bibitem{OG}
Hans~Christian \"Ottinger and Miroslav Grmela.
\newblock Dynamics and thermodynamics of complex fluids. {II}. {I}llustrations
  of a general formalism.
\newblock {\em Phys. Rev. E}, 56:6633--6655, Dec 1997.

\bibitem{HCO}
H.C. {\"O}ttinger.
\newblock {\em Beyond Equilibrium Thermodynamics}.
\newblock Wiley, 2005.

\bibitem{PhysicaD-2015}
M.~Pavelka, V.~Klika, O.~Esen, and M.~Grmela.
\newblock A hierarchy of {P}oisson brackets.
\newblock {\em Physica D}, submitted, 2015.

\bibitem{PRE2014}
Michal Pavelka, V\'aclav Klika, and Miroslav Grmela.
\newblock Time reversal in nonequilibrium thermodynamics.
\newblock {\em Phys. Rev. E}, 90:062131, Dec 2014.

\bibitem{Pavelka-IJES}
Michal Pavelka, Franti{\v s}ek Mar{\v s}{\' i}k, and V{\' a}clav Klika.
\newblock Consistent theory of mixtures on different levels of description.
\newblock {\em International Journal of Engineering Science}, 78(0):192 -- 217,
  2014.

\bibitem{ratiu1980motion}
Tudor~S Ratiu.
\newblock The motion of the free $ n $-dimensional rigid body.
\newblock {\em Indiana Univ. Math. J.}, 29(CAG-ARTICLE-1980-001):609--629,
  1980.

\bibitem{smale1970topology}
Steve Smale.
\newblock Topology and mechanics. i.
\newblock {\em Inventiones mathematicae}, 10(4):305--331, 1970.

\bibitem{Spencer-Kaufman}
RG~Spencer and AN~Kaufman.
\newblock Hamiltonian-structure of 2-fluid plasma dynamics.
\newblock {\em Physical Review A}, 25(4):2437--2439, 1981.

\bibitem{spivak1981comprehensive}
Michael Spivak.
\newblock comprehensive introduction to differential geometry. vol. i-iv.
\newblock 1981.

\bibitem{suhubi2013exterior}
Erdogan Suhubi.
\newblock {\em Exterior Analysis: Using Applications of Differential Forms}.
\newblock Elsevier, 2013.

\bibitem{Take81}
M.~Takeuchi.
\newblock Matched pairs of groups and bismash products of {H}opf algebras.
\newblock {\em Comm. Algebra}, 9(8):841--882, 1981.

\bibitem{Tu77}
W.M. Tulczyjew.
\newblock The {L}egendre transformation.
\newblock {\em Annales de l'IHP Physique théorique}, 27(1):101--114, 1977.

\bibitem{vaisman2012lectures}
Izu Vaisman.
\newblock {\em Lectures on the geometry of Poisson manifolds}, volume 118.
\newblock Birkh{\"a}user, 2012.

\bibitem{van2004port}
AJ~Van~der Schaft.
\newblock {\em Port-Hamiltonian systems: network modeling and control of
  nonlinear physical systems}.
\newblock Springer, 2004.

\bibitem{weinstein1983local}
Alan Weinstein.
\newblock The local structure of poisson manifolds.
\newblock {\em Journal of differential geometry}, 18(3):523--557, 1983.

\end{thebibliography}

\end{document}